\documentclass[referee, sn-mathphys-num]{sn-jnl}

\usepackage{amssymb}
\usepackage{amsfonts}
\usepackage[utf8]{inputenc}
\usepackage{graphicx}
\usepackage{dcolumn}
\usepackage{bm}
\usepackage{xcolor}
\usepackage{enumitem}
\usepackage[version=4]{mhchem}
\usepackage{microtype} 
\usepackage{comment}
\usepackage{booktabs}
\usepackage{multirow}
\usepackage{setspace}
\usepackage{makecell}
\usepackage{xr-hyper}
\usepackage{hyperref}
\usepackage{cleveref}
\usepackage{docmute}
\usepackage{listings}
\begin{document}

\title[SevenNet-Polar for MultiTask Prediction]{SevenNet-Polar for MultiTask Prediction of Energy, Forces, Stress, and Born Effective Charges: Development and Application to \ce{ZrO2}, \ce{Li3PO4}, and Perovskites}

\author*[1,2]{\fnm{Anh Khoa Augustin} \sur{Lu}}\email{LU.Augustin@nims.go.jp}
\author[1]{\fnm{Shungo} \sur{Arai}}
\author[3]{\fnm{Yutack} \sur{Park}}
\author[3]{\fnm{Seungwu} \sur{Han}}
\author[2,4,5]{\fnm{Tsuyoshi} \sur{Miyazaki}}
\author*[1]{\fnm{Satoshi} \sur{Watanabe}}\email{watanabe@cello.t.u-tokyo.ac.jp}

\affil[1]{\orgdiv{Department of Materials Engineering}, \orgname{The University of Tokyo}, \city{Tokyo}, \postcode{113-8656}, \country{Japan}}
\affil[2]{\orgdiv{Research Center for Materials Nanoarchitectonics (MANA)}, \orgname{National Institute for Materials Science (NIMS)}, \city{Tsukuba}, \state{Ibaraki}, \postcode{305-0044}, \country{Japan}}
\affil[3]{\orgdiv{Department of Materials Science and Engineering}, \orgname{Seoul National University}, \city{Seoul}, \postcode{08826}, \country{Korea}}
\affil[4]{\orgdiv{Graduate School of Engineering}, \orgname{Nagoya University}, \city{Nagoya}, \state{Aichi}, \postcode{464-8603}, \country{Japan}}
\affil[5]{\orgdiv{Master’s/Doctoral Program in Life Science Innovation}, \orgname{University of Tsukuba}, \city{Tsukuba}, \state{Ibaraki}, \postcode{305-8577}, \country{Japan}}

\abstract{
    Accurate prediction of the Born effective charge (BEC) tensor is crucial for modeling materials under electric fields but remains computationally expensive. To bridge this gap, we present \texttt{SevenNet-Polar}, an equivariant graph neural network framework based on the \texttt{SevenNet} architecture for fast and accurate BEC predictions. Our BEC-only predictors can achieve an RMSE as low as 0.0043~e on \ce{ZrO2}, \ce{Li3PO4}, and perovskites, despite the presence of high-temperature (up to 2,000 K) and defect-laden training data. Our all-in-one multitask models for predicting energy, forces, stress, and BEC in \ce{ZrO2} and \ce{Li3PO4} achieve high accuracy with an RMSE of 1.0~meV/atom for energy, 12 meV/\AA~for forces, 0.05~GPa for stress, and 0.0029~e for BEC. BEC accuracy is not degraded by multitask training. Scaling analysis reveals distinct exponents for diagonal and off-diagonal BEC components, both of which exhibit less favorable scaling than energy, force and stress errors. SevenNet-Polar generalizes robustly when tested on scenarios containing structural environments absent from the training set, such as along nudged elastic band (NEB) trajectories or grain boundaries in \ce{ZrO2}. Accelerated by FlashTP, SevenNet-Polar enables simulations containing up to 1.5~million atoms on multi-GPU supercomputers and up to approximately 15,000 atoms on a single consumer-grade GPU. This makes charge-aware molecular dynamics simulations under electric fields more accessible.
}

\maketitle

\section{Introduction}

    The development of machine learning interatomic potentials (MLIPs) has transformed computational materials science by enabling models that reproduce the results of first-principles simulations at a fraction of the cost, with substantially higher accuracy than empirical potentials. While earlier works have probed the ability of neural networks (NN) to fit potential energy surfaces (PES)~\cite{Blank_1995,Lorenz_2004}, these models were typically limited to a fixed number of atoms and specific geometries, thus making large-scale MD calculations arduous. The introduction in 2007 of the High-Dimensional Neural Network Potential (HDNNP) by Behler~\cite{Behler_2007} using atom-centered symmetry functions (ACSFs) allowed MLIPs to be applied to systems of any size and triggered an unprecedented growth in the field. 

    More recently, graph neural networks (GNNs) with message passing have emerged as a promising architecture. By representing atomic structures as graphs where atoms serve as nodes and interatomic distances as edges, message passing neural networks (MPNNs~\cite{Gilmer_2017}) allow information to dynamically propagate, thereby extending the effective receptive field far beyond the local cutoff. Notable architectures include SchNet~\cite{Schutt_2017}, Neural Equivariant Interatomic Potentials (NequIP)~\cite{Batzner_2022}, and MACE~\cite{Batatia_2022}. Equivariant GNNs have become the backbone of universal machine learning potentials (uMLIPs), which can describe over 80 chemical elements, such as the MACE Foundation Model (MACE-MP-0~\cite{Batatia_2025}), SevenNet-0~\cite{Park_2024}, SevenNet-Omni~\cite{Kim_2026}, or PET-OAM-XL~\cite{Mazitov_2025}, enabling simulations of a large variety of materials.

    In the present work, we extend the SevenNet architecture to enable prediction of the BEC tensor. As graph neural networks are shown to be able to mimic long-range interaction via message passing~\cite{Kang_2025}, we set four convolution layers to extend the receptive field (though recent universal models have pushed that number to up to 12~\cite{Kim_2026}). First, we trained SevenNet-Polar specialized  models for BEC prediction (SevenNet-PS) on a data set comprising \ce{ZrO2}, \ce{Li3PO4}, and perovskite structures. We propose three model sizes (SevenNet-PS-S for small, SevenNet-PS-M for medium and SevenNet-PS-L for large), achieving remarkable accuracy (0.0120, 0.0063 and 0.0043~$e$ for BEC for SevenNet-PS-S/M/L, respectively). We carefully studied the power law between the root mean squared errors (RMSE) and the size of the training data ($RMSE \propto N^{-\alpha}$) and showed that a difference exists between diagonal ($\alpha\approx 0.38$) and off-diagonal terms ($\alpha\approx0.32$). Then, we trained unified multitask models (SevenNet-PM) capable of simultaneously predicting energy, forces, stress, and the BEC tensor, again with three model sizes (SevenNet-PM-S for small, SevenNet-PM-M for medium and SevenNet-PM-L for Large). The accuracy was evaluated on a data set containing \ce{ZrO2} and \ce{Li3PO4} structures, with defects and high-temperature (up to 2,000~K) configurations. We obtained excellent accuracy with an RMSE of 1.0~meV/atom on the energy, 11.9~meV/\AA~on the forces, 0.05~GPa on the stress tensor, and 0.0029~$e$ on the BEC tensor. This level of accuracy could be obtained by setting an extended receptive field of 30~\AA~(five layers, cutoff of 6~\AA), and using higher-order representations ($l_{\text{max}}=4$), leading to models with around 5-7 million parameters. The small models, on the other hand, have only 500,000-700,000 parameters, being more suitable for fast molecular dynamics simulations. Power scaling law analysis for the different quantities predicted showed a hierarchy of difficulty from the fitted exponent, with the energy ($\alpha\approx 0.73$), stress tensor ($\alpha\approx 0.72$), forces ($\alpha\approx 0.53$), then BEC diagonal components ($\alpha\approx 0.39$), and finally BEC off-diagonal components ($\alpha\approx 0.32$). It is noteworthy that multitask training is not detrimental to the accuracy of BEC predictions. 
    
    To manage the high computational cost associated with deep equivariant networks, our integration is compatible with FlashTP~\cite{Lee_2025}, a high-performance library that replaces specific computationally expensive tensor-product operations within standard frameworks like \texttt{e3nn}~\cite{Geiger_2022}, enabling significantly faster model training and scaling molecular dynamics simulations to much larger systems while drastically reducing the memory usage. Integrated with the Atomic Simulation Environment (ASE)~\cite{Larsen_2017} and the Large-scale Atomic/Molecular Massively Parallel Simulator (\texttt{LAMMPS})~\cite{Plimpton_1995}, the model facilitates seamless execution of molecular dynamics and nudged elastic band (NEB) calculations, achieving speeds of 1-10 ns/day on a single GPU. 

    Finally, we validate the predictive capabilities of the model on structures not included in the training set with two test cases. Firstly, a nudged elastic band (NEB) calculation pathway of oxygen migration in defect-laden \ce{ZrO2} with an oxygen vacancy, which contains several out-of-equilibrium structures. Secondly, a $\Sigma 5 (310)$ grain boundary in cubic-\ce{ZrO2}~\cite{Arai_2025}, which has a different number of atoms and different local environments compared to the structures included in the training set. 

    The methodology developed in the present work establishes a comprehensive framework for describing materials under applied electric fields. The models developed in this work cover up to ten chemical elements, indicating that the approach can be extended to chemically diverse data sets. 

\section{Results}


        \subsection{Scaling law for BEC prediction for the \texorpdfstring{\ce{ZrO2}}{ZrO2} data set}
            
\begin{figure}
                \centering
                \includegraphics[width=\columnwidth]{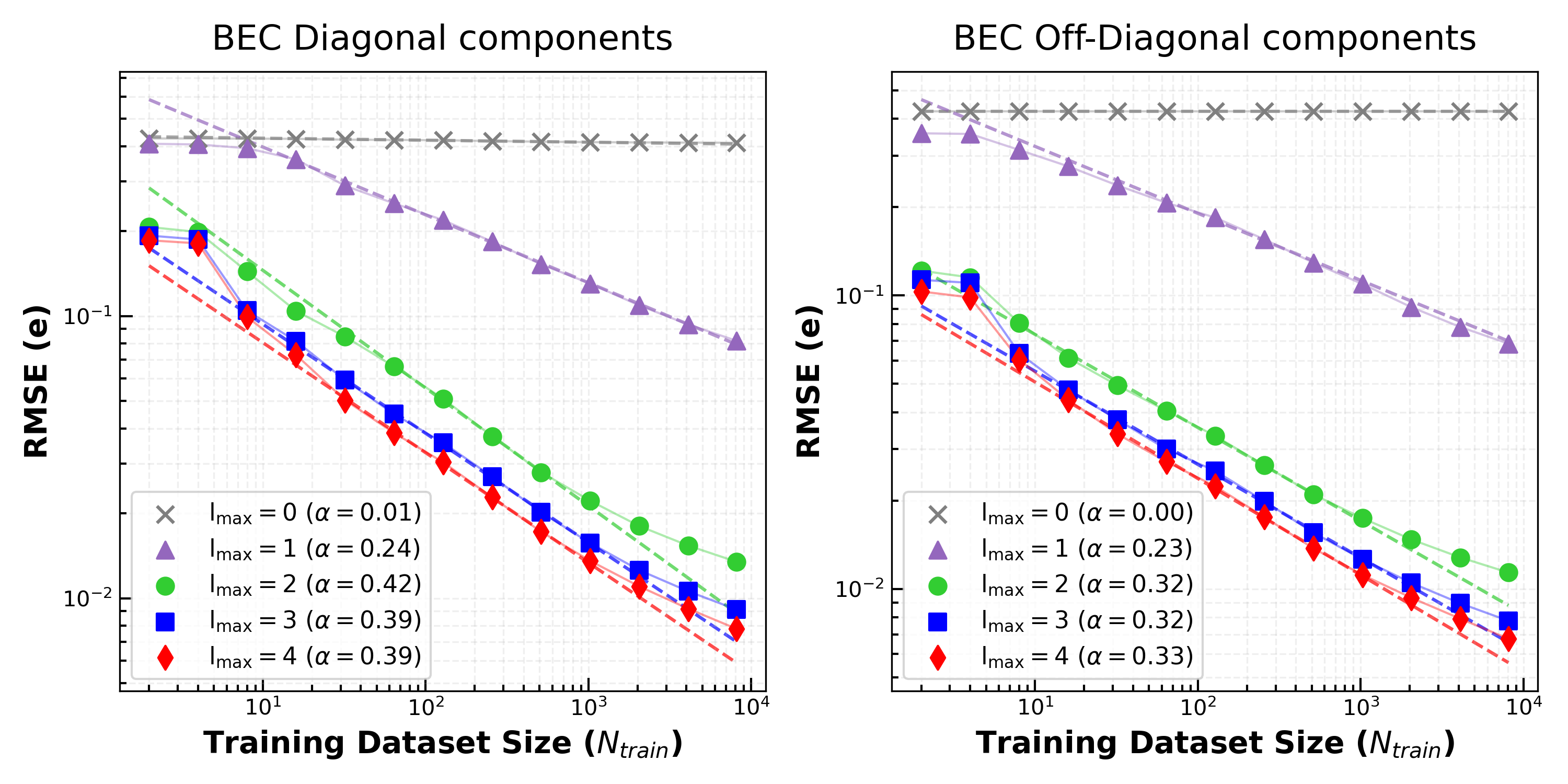}
                \caption{Power-law scaling of SevenNet-PS models on BEC diagonal (left) and off-diagonal (right) component prediction errors on a test set for $l_{\text{max}}$ ranging from 0 to 4. The scaling exponent $\alpha$ is derived from fits performed between $N_{\text{train}} \in [64, 512]$.}
                \label{fig:bec_power_law}
            \end{figure}

            To assess the accuracy of the newly implemented BEC prediction layers, we first trained a SevenNet-PS model solely on the BEC tensor using data from the \ce{ZrO2} data set. To evaluate the scaling law with respect to the maximum complexity of the spherical harmonics $l_{\text{max}}$, we split our data set as 80\% train, 10\% validation and 10\% test sets and train on part of this training set (using 2, 4, 8,... 4096, and 8082 structures), while maintaining the same validation and test sets. The evolution of the RMSE for the diagonal and off-diagonal BEC components is presented in \autoref{fig:bec_power_law} as log-log plot for $l_{\text{max}}$ ranging from 0 to 4. A power-law fit, $\text{RMSE} \propto N_{train}^{-\alpha}$, is applied to evaluate the scaling exponent $\alpha$. As expected for the prediction of tensorial properties, the scalar-only model ($l_{\text{max}}=0$) completely fails to learn the BECs and the exponent remains near zero, as the error does not  decrease despite the addition of training data. Conversely, for $l_{\text{max}} \geq 1$, both the diagonal and off-diagonal errors exhibit strict power-law scaling before saturating near $N_{\text{train}} \approx 4000-8000$. It is noteworthy that the off-diagonal components follow a different power law, decreasing more slowly than the diagonal elements as the data set expands. Increasing the angular resolution systematically steepens the learning curve: for the diagonal (resp. off-diagonal) components, the scaling exponent $\alpha$ improves from 0.24 (0.23) at $l_{\text{max}}=1$, to 0.42 (0.32) at $l_{\text{max}}=2$, 0.39 (0.32) at $l_{\text{max}}=3$, and 0.39 (0.33) at $l_{\text{max}}=4$. This added tensorial complexity at each step significantly lowers the asymptotic RMSE, demonstrating that $l_{\text{max}} = 3$ is necessary to achieve accuracies below 0.01~$e$ for \ce{ZrO2}. Increasing $l_{\text{max}}$ to 4 slightly decreases the RMSE, but more than doubles the cost as the time per epoch increases with $l_{\text{max}}$ from 100, 149, 172, 261 and 598 s/epoch on an NVIDIA GeForce RTX 4090 as $l_{\text{max}}$ increases from 0 to 4. Therefore $l_{\text{max}} = 3$ is considered the sweet spot for high-precision models (SevenNet-PS-M and SevenNet-PM-M), while $l_{\text{max}} = 2$ is considered a good compromise for high-speed models (SevenNet-PS-S and SevenNet-PM-S). Nevertheless, for applications that would require the highest possible accuracy, we propose SevenNet-PS-L and SevenNet-PM-L with $l_{\text{max}} = 4$ and an extended receptive field with 5 layers.
        
        \subsection{SevenNet-PS: Specialized models for BEC prediction}

\begin{figure*}[htbp]
                \centering
                \includegraphics[width=0.8\linewidth]{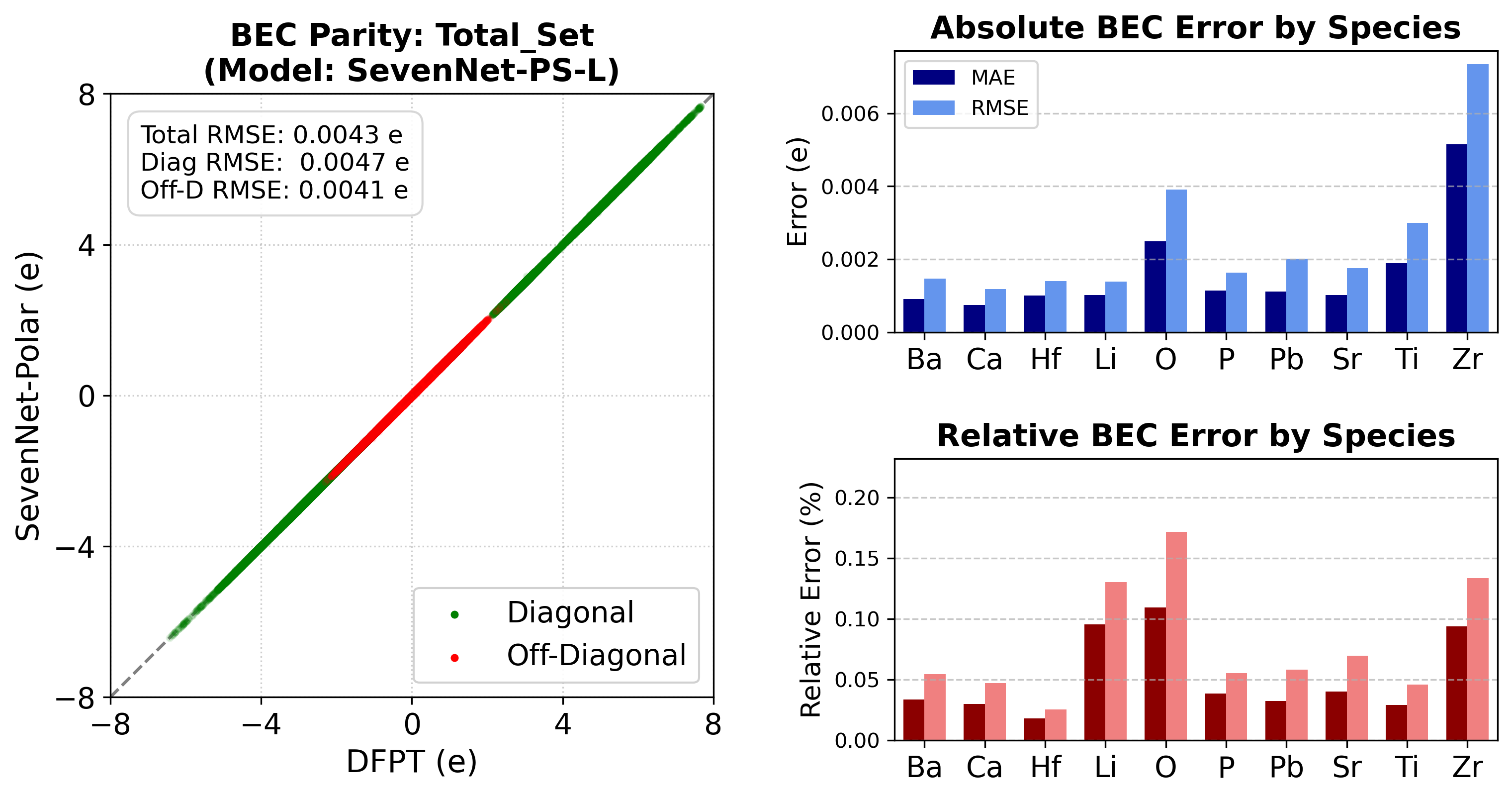}
                \caption{Evaluation of the accuracy of SevenNet-PS-M on the combined multi-material test set (\ce{ZrO2}, \ce{Li3PO4}, and perovskite). (left) Parity plots comparing BEC tensor components predicted by SevenNet-Polar and DFPT data. (right) Distribution of prediction errors for each of the ten chemical species.}
                \label{fig:specialized_on_triple_test}
            \end{figure*}

            \autoref{fig:specialized_on_triple_test} presents the results for the SevenNet-PS-L (SevenNet-Polar Specialized Large) model trained on \ce{ZrO2}, \ce{Li3PO4} and perovskites data sets. A root mean squared error (RMSE) of 0.0047 e was obtained on the diagonal components of the BEC tensor while a lower RMSE of 0.0041 e was obtained for the off-diagonal components, showing that both types of components are properly fitted. The total RMSE is 0.0043 e. A detailed analysis on each atomic species shows RMSE values of 0.005 e or below for all species except Zr. However, normalizing the error with respect to the average value shows that the relative error is comparable to other species (below 0.1 \%). The relative error on oxygen being slightly higher can be related to the fact that these atoms appear in all the data sets and are therefore subject to a greater variation of atomic environments (\ce{ZrO2}, \ce{Li3PO4} and perovskites). These values are nevertheless much lower than in previous works by Shimizu \cite{Shimizu_2023} and Kutana \cite{Kutana_2025}, as shown in \autoref{tab:comparison_rmse_bec}. We also propose a smaller version of our model, SevenNet-PS-M (resp. SevenNet-PS-S) where the number of channels is 64 (resp. 32), the number of layers is set to $L=4$, and $l_{\text{max}}$ is set to 3 (resp. 2). This divides the number of parameters by two (resp. ten), while reaching an RMSE of 0.063 e (resp. 0.0120 e). Details can be found in \cref{sec:sevennet_polar_small_models} in the Supplementary Information. As we tracked the origin of each structure in the test set, we also analyzed the performance of our model on each data set, as shown in the Supplementary Information. 
    
            On \ce{ZrO2}, we obtained RMSE values of 0.0063 e, 0.0084 e and 0.0152 e for the SevenNet-PS-L, SevenNet-PS-M and SevenNet-PS-S models, which is lower than the values reported by Kutana \cite{Kutana_2025} and Lu \cite{Lu_2026}. It is noteworthy that SevenNet-PS-S has a similar number of parameters as the \texttt{equivar} BM1 model, but shows a reduction of the RMSE by a factor 3. On \ce{Li3PO4}, the improvement is more pronounced, as SevenNet-PS-S achieved a reduction of the RMSE by a factor 4 compared to the previous models by Kutana \cite{Kutana_2025} and Shimizu \cite{Shimizu_2023}. Here the RMSE values were 0.0015 e, 0.0028 e and 0.0075 e, respectively.  Finally, on the perovskite test set, excellent RMSE values of 0.0017 e,  0.0022 e and 0.0045 e were obtained. Across all three data sets, SevenNet-Polar achieves a reduction of the RMSE by 85\% to 95\% compared to previous models. In terms of mean absolute error (MAE), SevenNet-PS-L has a value of 0.0026 e, which is slightly smaller than values reported by Falletta et al. with \texttt{allegro-pol} \cite{Falletta_2025}, which was trained on \ce{SiO2} and \ce{BaTiO3}. However, an important distinction should be made as our data sets include high-temperature structures and defect-laden structures, which are typically detrimental to the accuracy. 
            
            The good accuracy among 10 different atomic species shows that the SevenNet-Polar architecture can accommodate a large number of elements, and could ultimately be extended to become a universal model, if enough data points can be obtained.
        
            \begin{table*}[htbp]
                \caption{\label{tab:comparison_rmse_bec} Comparison of model architectures and predictive performance (RMSE) for Born effective charges on test sets, also separated by type of data (\ce{ZrO2} pristine, \ce{ZrO2} including defects, \ce{Li3PO4} and perovskites). Architectural hyperparameters include the cutoff radius ($r_c$), number of message-passing layers ($L$), feature channels ($C$), and maximum angular momentum ($l_{\text{max}}$). The proposed SevenNet-Polar models are compared with recent literature in terms of RMSE. Parity plots can be found in \cref{sec:sevennet_ps_m_on_each_data_set} of the Supplementary information.}
                \footnotesize 
                \begin{tabular}{@{}ll cccc c c@{}}
                    \toprule
                    \textbf{Model} & \textbf{Data set} & $r_c$ & $L$ & $C$ & $l_{\text{max}}$ & BEC$_{\text{total}}$ & Parameters \\
                     & & (\AA) & & & & ($e$) & \\
                    \midrule
                    SevenNet-PS-L& \ce{ZrO2}+\ce{Li3PO4}+perovskites & 6.0 & 5 & 64 & 4 & 0.0043 & 7,578,504 \\
                    SevenNet-PS-M& \ce{ZrO2}+\ce{Li3PO4}+perovskites & 6.0 & 4 & 64 & 3 & 0.0063 & 3,546,952 \\
                    SevenNet-PS-S & \ce{ZrO2}+\ce{Li3PO4}+perovskites & 6.0 & 4 & 32 & 2 & 0.0120 & 609,992 \\
                    \midrule
                    SevenNet-PS-L & \ce{ZrO2} & 6.0 & 5 & 64 & 4 & 0.0059 & 7,578,504 \\
                    SevenNet-PS-M & \ce{ZrO2} & 6.0 & 4 & 64 & 3 & 0.0084 & 3,546,952 \\
                    SevenNet-PS-S & \ce{ZrO2}   & 6.0 & 4 & 32 & 2 & 0.0152 & 609,992 \\
                    Lu et al. \cite{Lu_2026} & \ce{ZrO2} & 7.0 & -- & -- & -- & 0.0936 & 4,682 \\ 
                    \texttt{equivar} BM1 \cite{Kutana_2025} & \ce{ZrO2} & 3.0 & 6 & 64 & 2 & 0.0510 & 510,000 \\  
                    \texttt{equivar} BM2 \cite{Kutana_2025} & \ce{ZrO2} & 3.0 & 6 & 32 & 2 & 0.0569 & 140,000 \\
                    \texttt{equivar} Spec. \cite{Kutana_2025} & \ce{ZrO2} & 3.0 & 6 & 32 & 2 & 0.0493 & 131,000 \\   
                    \midrule
                    SevenNet-PS-L & \ce{Li3PO4} & 6.0 & 5 & 64 & 4 & 0.0015 & 7,578,504 \\
                    SevenNet-PS-M & \ce{Li3PO4} & 6.0 & 4 & 64 & 3 & 0.0028 & 3,546,952 \\
                    SevenNet-PS-S & \ce{Li3PO4} & 6.0 & 4 & 32 & 2 & 0.0075 & 609,992 \\                
                    Shimizu et al. \cite{Shimizu_2023} & \ce{Li3PO4} & 7.0 & -- & -- & -- & 0.0376 & 5,793 \\
                    \texttt{equivar} BM1 \cite{Kutana_2025} & \ce{Li3PO4} & 3.0 & 6 & 64 & 2 & 0.0319 & 510,000 \\
                    \texttt{equivar} BM2 \cite{Kutana_2025} & \ce{Li3PO4} & 3.0 & 6 & 32 & 2 & 0.0364 & 140,000 \\
                    \texttt{equivar} Spec. \cite{Kutana_2025} & \ce{Li3PO4} & 3.0 & 6 & 32 & 2 & 0.0325 & 131,000 \\
                    \midrule
                    SevenNet-PS-L & perovskites & 6.0 & 5 & 64 & 4 & 0.0017 & 7,578,504 \\                
                    SevenNet-PS-M & perovskites & 6.0 & 4 & 64 & 3 & 0.0022 & 3,546,952 \\                
                    SevenNet-PS-S & perovskites & 6.0 & 4 & 32 & 2 & 0.0045 & 609,992 \\                  
                    \texttt{equivar} BM1 \cite{Kutana_2025} & perovskites & 3.0 & 6 & 64 & 2 & 0.0300 & 510,000 \\
                    \texttt{equivar} BM2 \cite{Kutana_2025} & perovskites & 3.0 & 6 & 32 & 2 & 0.0377 & 140,000 \\
                    \texttt{equivar} Spec. \cite{Kutana_2025} & perovskites & 3.0 & 6 & 32 & 2 & 0.0135 & 131,000 \\
                    \botrule
                \end{tabular}
            \end{table*}
        
    \subsection{SevenNet-PM: Multitask learning}

        \subsubsection{Scaling Laws for multitask learning}
        
            \begin{figure*}[htbp]
                \centering
                \includegraphics[width=0.9\linewidth]{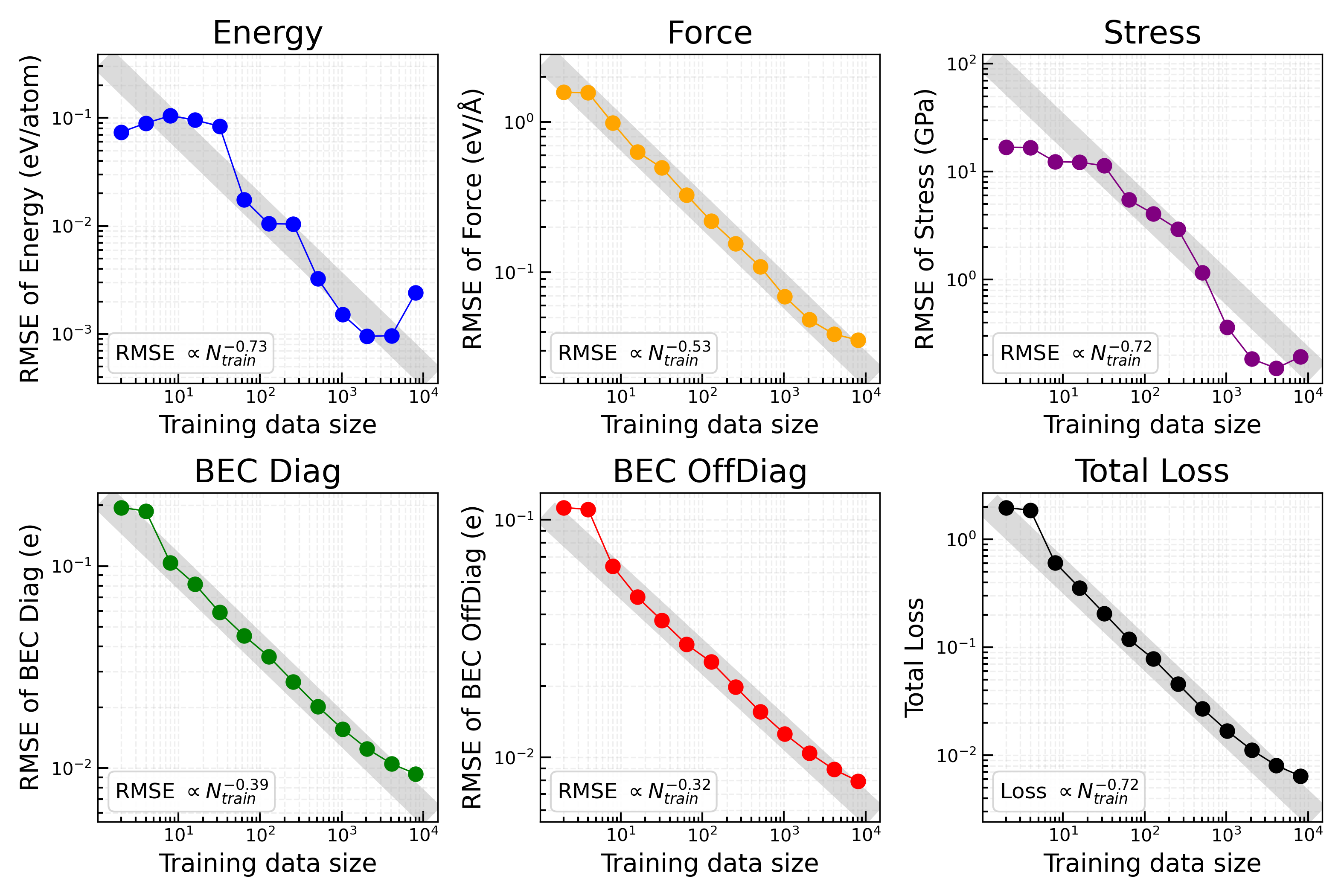}
                \caption{\label{fig:Scaling_law_multitask} Scaling behavior of SevenNet-PM-M on the \ce{ZrO2} data set. Root-mean-square errors (RMSE) for energy, forces, stress, and Born effective charge (BEC) components, alongside the total loss, are plotted against the training data set size ($N_{\text{train}}$) on a log-log scale. The thick gray bands represent power-law fits ($RMSE \propto N_{\text{train}}^{-\alpha}$) calculated for the intermediate data regime ($64 \le N_{\text{train}} \le 512$). The extracted scaling exponents ($\alpha$) are provided in the insets.}
            \end{figure*}
        
            The present implementation of SevenNet-Polar enables seamless multitask training on multiple quantities, enabling simultaneous prediction of energy, forces, stress, and Born effective charges with a single model, eliminating the need to combine multiple models as was done in previous works \cite{Shimizu_2023,Kutana_2025,Lu_2026}. To systematically evaluate the data efficiency and learning capacity of our multitask SevenNet-PM architecture, we conducted a scaling law analysis using the pristine and defective \ce{ZrO2} data set (\autoref{fig:Scaling_law_multitask}). The model was trained on progressively larger subsets ranging from 20 to 8,082 structures sampled from the full training set. Model performance is reported as the test set RMSE evaluated at the epoch that minimized the validation loss. 
            
            As illustrated in \autoref{fig:Scaling_law_multitask}, all learned quantities exhibit robust power-law decay ($\text{RMSE} \propto N_{\text{train}}^{-\alpha}$). Importantly, multitask training is not detrimental to the accuracy of the BEC predictions, though each physical property converges at a distinctly different rate. The macroscopic properties, energy and stress, demonstrate the steepest scaling exponents of $\alpha = 0.73$ and $0.72$, respectively. On the other hand, the atomic forces scale with an exponent of $0.53$. 
            
            The BEC tensor components exhibit shallower scaling exponents, with the diagonal and off-diagonal terms decaying at $\alpha = 0.39$ and $0.32$, respectively. This slower convergence highlights the inherent difficulty in learning highly anisotropic, non-local dielectric responses compared to localized structural targets, corroborating the tensor modeling complexities discussed by Kutana et al.~\cite{Kutana_2025}. Finally, the slight flattening observed at the extreme right of the curves indicates the onset of data saturation, suggesting that the full training set of $\sim$8,000 structures is fairly well sized to capture the underlying physics before experiencing diminishing returns.

        \subsubsection{SevenNet-PM-L, SevenNet-PM-M, and SevenNet-PM-S: Multitask models}

\begin{figure*}[htpb]
                \centering
                \includegraphics[width=\linewidth]{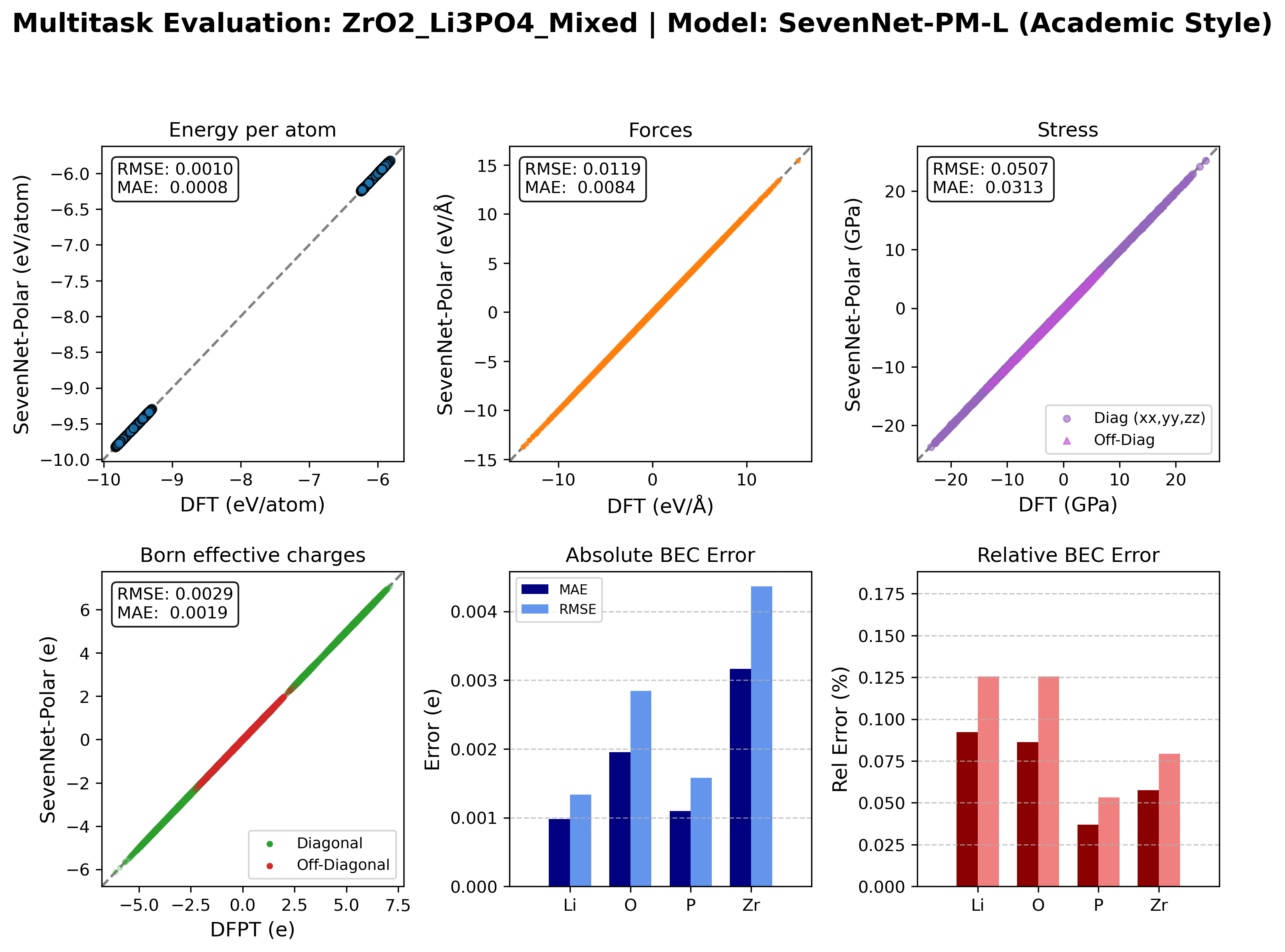}
                \caption{Parity plots for total energy, forces, stress tensor and BEC tensor for SevenNet-PM-L. This model has been trained on the \ce{ZrO2} and \ce{Li3PO4} data sets.}
                \label{fig:multitask_best}
            \end{figure*}

            For the SevenNet-PM models, we focused on \ce{ZrO2} and \ce{Li3PO4} data sets to maintain consistent PBE-level energy, forces and stress. \autoref{fig:multitask_best} presents the parity plots of SevenNet-PM-L trained on \ce{ZrO2} and \ce{Li3PO4} data sets, as well as the error on each species. This model, on the test set, shows an RMSE of 1.0 meV/atom on the energy, 11.9 meV/\AA~ on the forces, 0.05 GPa on the stress tensor and 0.0029 e on the BEC tensor (0.0031 e for diagonal components and 0.0028 e for off-diagonal elements). It should be noted that the BEC RMSE is lower than that of the BEC-only model. This originates from the absence of the perovskite structures, whose number is small, while presenting a larger variety of atomic species. The accuracy of this model makes it suitable for use in various simulations such as molecular dynamics simulations, nudged-elastic-band (NEB) calculations or phonon band structure calculation.
        
            \autoref{tab:comparison_rmse} reports the performances of different models and provides a comparison with the literature. We also included standalone models that were trained on a single data set. The \ce{ZrO2} pristine data set is a subset of the \ce{ZrO2} data set containing only 3,977 structures out of the 10,103 original structures, where structures with defects are excluded. As for the specialized SevenNet-PS models, we also propose a lighter version of our multitask SevenNet-PM models, SevenNet-PM-M and SevenNet-PM-S, defined with the same hyperparameters as for the SevenNet-PS series, with half or a tenth of the number of parameters of SevenNet-PM-L, respectively. Compared to previous works, the SevenNet-PM-S (trained on multiple data sets) achieves better accuracy than previous works such as the approaches by Shimizu \cite{Shimizu_2023} and Lu \cite{Lu_2026}, who used High-Dimensional Neural Network Potentials (HDNNPs) neural networks \cite{Behler_2007} and vector atomic fingerprints (VAF) \cite{Li_2017}, or \texttt{equivar}\cite{Kutana_2025}, who relied on equivariant graph neural networks. Detailed plots of the errors can be found in the Supplementary Information. Our best models outperform all these models on the BEC metrics, while providing similar accuracy to the \texttt{allegro-pol} model with the lowest error. It should be noted that \texttt{allegro-pol} is trained on a single composition \ce{BaTiO3} with a data set of low-temperature structures, while our model was trained on two data sets for multitask learning or three data sets for BEC-only learning, including high-temperature structures, many of which contained point defects. This shows that our architecture is well suited for training on multiple data sets, which is promising for potential future development of a universal model, although this remains challenging because DFPT data sets are scarce.
                    
            \begin{table*}[ht]
            \caption{\label{tab:comparison_rmse} Comparison of model architectures and predictive performance (RMSE) on test sets. Architectural hyperparameters include the cutoff radius ($r_c$), number of message-passing layers ($L$), feature channels ($C$), and maximum angular momentum ($l_{\text{max}}$). The proposed SevenNet-based models are benchmarked against recent literature reporting RMSE. Dashes (--) indicate properties not targeted during training.}
            \footnotesize 
            \begin{tabular}{@{}ll cccc ccc cc@{}}
            \toprule
            \textbf{Model} & \textbf{Data set} & $r_c$ & $L$ & $C$ & $l_{\text{max}}$ & Energy & Force & Stress & BEC & Parameters \\
             & & (\AA) & & & & (meV/atom) & (meV/\AA) & (GPa) & ($e$) & \\
            \midrule
            SevenNet-PM-L  & \ce{ZrO2}+\ce{Li3PO4} & 6.0 & 5 & 64 & 4 & 1.0 & 11.9 & 0.05 & 0.0029 & 5,833,224 \\
            SevenNet-PM-M  & \ce{ZrO2}+\ce{Li3PO4} & 6.0 & 4 & 64 & 3 & 0.8 & 18.0 & 0.05 & 0.0063 & 2,538,952 \\
            SevenNet-PM-S & \ce{ZrO2}+\ce{Li3PO4} & 6.0 & 4 & 32 & 2 & 1.7 & 36.9 & 0.09 & 0.0121 & 431,624 \\        
            \midrule
            SevenNet-PM-L  & \ce{ZrO2} & 6.0 & 5 & 64 & 4 & 1.6 & 12.7 & 0.07 & 0.0038 & 5,833,224 \\
            SevenNet-PM-M  & \ce{ZrO2} & 6.0 & 4 & 64 & 3 & 1.2 & 19.3 & 0.05 & 0.0084 & 2,202,952 \\
            SevenNet-PM-S & \ce{ZrO2} & 6.0 & 4 & 32 & 2 & 2.7 & 40.5 & 0.12 & 0.0151 & 431,624 \\
            Lu et al.\cite{Lu_2026} & \ce{ZrO2} & 7.0 & -- & -- & -- & 5.1 & 99.0 &  -- & 0.0936 & 4,682 \\
            \midrule
            SevenNet-PM-L  & \ce{Li3PO4} & 6.0 & 5 & 64 & 4 & 0.4 & 11.4 & 0.03 & 0.0014 & 5,833,224 \\ 
            SevenNet-PM-M  & \ce{Li3PO4} & 6.0 & 4 & 64 & 3 & 0.4 & 16.4 & 0.04 & 0.0028 & 2,370,952 \\ 
            SevenNet-PM-S & \ce{Li3PO4} & 6.0 & 4 & 32 & 2 & 0.8 & 32.7 & 0.08 & 0.0075 & 431,624 \\        
            Shimizu et al.\cite{Shimizu_2023} & \ce{Li3PO4} & 7.0 & -- & -- & -- & 2.9 & 87.9 & -- & 0.0376 & 5,793 \\
            \botrule
            \end{tabular}
            \end{table*}

        \subsubsection{Hyperparameters and performances}

            An ablation study of the hyperparameters on the \ce{ZrO2} data set is presented in \autoref{tab:ablation_combined} to analyze the importance of each hyperparameter during multitask learning. We observe that the maximum angular momentum ($l_{\text{max}}$) and the message-passing depth (or number of layers) define the performance of our models, revealing the necessity of incorporating higher-order geometric features to properly learn the BEC tensors on one hand, and expanding the receptive field range on the other hand. Furthermore, we observe that while a short cutoff radius of $r_c = 3.0$~\AA{} yields a reasonable BEC accuracy (RMSE $\approx 0.02~e$), consistent with the \texttt{equivar} model~\cite{Kutana_2025}, it results in sub-par performance for atomic forces and stress tensors compared to state-of-the-art standards. Systematically increasing this cutoff improves all target metrics. The timing is taken on a single NVIDIA A100 GPU.
            
            \begin{table*}[htpb]
                \centering
                \caption{Ablation studies for the neural network architecture hyper-parameters  on the \ce{ZrO2} data set. The reference model ($l_{\text{max}}=3$, $L=4$, $r_c=6.0$~\AA, $C=64$) optimally balances predictive accuracy for Born effective charges and hardware-accelerated execution time. Force and Stress RMSE are reported component-wise to ensure consistency with standard literature benchmarks. (* indicates that \texttt{FlashTP} could not be used)}
                \label{tab:ablation_combined}
                \resizebox{0.9\textwidth}{!}{%
                \begin{tabular}{lcccccc}
                \toprule
                \textbf{Parameter} & \textbf{Time/Ep (s)} & \textbf{E (meV/atom)} & \textbf{F (meV/\AA)} & \textbf{S (GPa)} & \textbf{BEC Diag ($e$)} & \textbf{BEC Off-Diag ($e$)} \\
                \midrule
                \multicolumn{7}{l}{\textbf{Maximum angular resolution ($l_{\text{max}}$)}} \\
                \midrule
                0 & 83 & 6.5 & 219.5 & 0.73 & 0.4114 & 0.4237 \\
                1 & 149 & 3.0 & 125.6 & 0.40 & 0.0803 & 0.0680 \\
                2 & 192 & 2.1 & 29.6 & 0.07 & 0.0134 & 0.0113 \\
                \textbf{3 (SevenNet-PM-M)} & \textbf{264} & \textbf{1.0} & \textbf{18.7} & \textbf{0.04} & \textbf{0.0091} & \textbf{0.0078} \\
                4 & 796 & 0.7 & 15.7 & 0.05 & 0.0078 & 0.0067 \\
                \midrule
                \multicolumn{7}{l}{\textbf{Message-passing depth ($L$)}} \\
                \midrule
                1 & 73 & 10.3 & 261.7 & 0.94 & 0.1337 & 0.1215 \\
                2 & 125 & 2.7 & 55.4 & 0.14 & 0.0378 & 0.0299 \\
                3 & 219 & 1.5 & 27.8 & 0.09 & 0.0142 & 0.0120 \\
                \textbf{4 (SevenNet-PM-M)} & \textbf{264} & \textbf{1.0} & \textbf{18.7} & \textbf{0.04} & \textbf{0.0091} & \textbf{0.0078} \\
                5 & 425 & 0.8 & 15.2 & 0.04 & 0.0072 & 0.0063 \\
                \midrule
                \multicolumn{7}{l}{\textbf{Distance cutoff ($r_c$ in \AA)}} \\
                \midrule
                3.0 & 227 & 3.9 & 53.1 & 0.18 & 0.0191 & 0.0160 \\
                4.0 & 211 & 1.2 & 23.4 & 0.09 & 0.0109 & 0.0094 \\
                5.0 & 233 & 1.3 & 19.6 & 0.05 & 0.0093 & 0.0080 \\
                \textbf{6.0 (SevenNet-PM-M)} & \textbf{264} & \textbf{1.0} & \textbf{18.7} & \textbf{0.04} & \textbf{0.0091} & \textbf{0.0078} \\
                7.0 & 448 & 0.8 & 18.7 & 0.05 & 0.0090 & 0.0077 \\
                \midrule
                \multicolumn{7}{l}{\textbf{Internal feature dimension (Channels, $C$)}} \\
                \midrule
                8 & 674* & 2.9 & 53.7 & 0.23 & 0.0185 & 0.0152 \\
                16 & 1074* & 1.4 & 35.0 & 0.14 & 0.0137 & 0.0115 \\
                32 & 222 & 0.8 & 24.5 & 0.06 & 0.0107 & 0.0090 \\
                \textbf{64 (SevenNet-PM-M)} & \textbf{264} & \textbf{1.0} & \textbf{18.7} & \textbf{0.04} & \textbf{0.0091} & \textbf{0.0078} \\
                128 & 555 & 2.0 & 19.3 & 0.17 & 0.0092 & 0.0078 \\
                \bottomrule
                \end{tabular}%
                }
            \end{table*}
            
            Although increasing $l_{\text{max}}$, the number of channels, and the message-passing depth inherently increases computational overhead, the integration of \texttt{FlashTP}~\cite{Lee_2025} dramatically mitigates these memory and time costs. This optimization enables the training of highly expressive, multi-million parameter equivariant models at a reasonable cost, even on a single consumer-grade NVIDIA GeForce RTX 4090 with 24~GB of VRAM. The use of \texttt{FlashTP} even allowed training models with $l_{\text{max}}=4$, though the time per epoch increased dramatically from 264 s/epoch to 796 s/epoch on an NVIDIA A100 GPU. An optimal balance between accuracy and computational efficiency is achieved with $r_c = 6.0$~\AA{} and $L=4$ message-passing layers, yielding an effective receptive field of 24~\AA. These hyperparameters serve as the foundation for our main SevenNet-PS-M and SevenNet-PM-M  models. By setting $l_{\text{max}}=3$ and feature channels $C=64$, SevenNet-PS-M and SevenNet-PM-M contain 3,546,952 and 2,538,952 parameters, respectively. The higher parameter count in SevenNet-PS-M arises from the expanded atomic embedding dictionary required for the larger number of chemical species present in the combined data set. For applications requiring high accuracy for single or few evaluations, we also propose models with higher complexity $l_{\text{max}}=4$, and $L=5$, labeled SevenNet-PS-L and SevenNet-PM-L, with 7,578,504 and 5,833,224 parameters, respectively.   
            
            For applications demanding faster inference, such as long molecular dynamics simulations, we also propose a lighter architecture with reduced complexity by setting $l_{\text{max}}=2$ and $C=32$, reducing the parameter count to 609,992 (SevenNet-PS-S) and 431,624 (SevenNet-PM-S). 
        
            \begin{figure}[htbp]
                \centering
                \includegraphics[width=\linewidth]{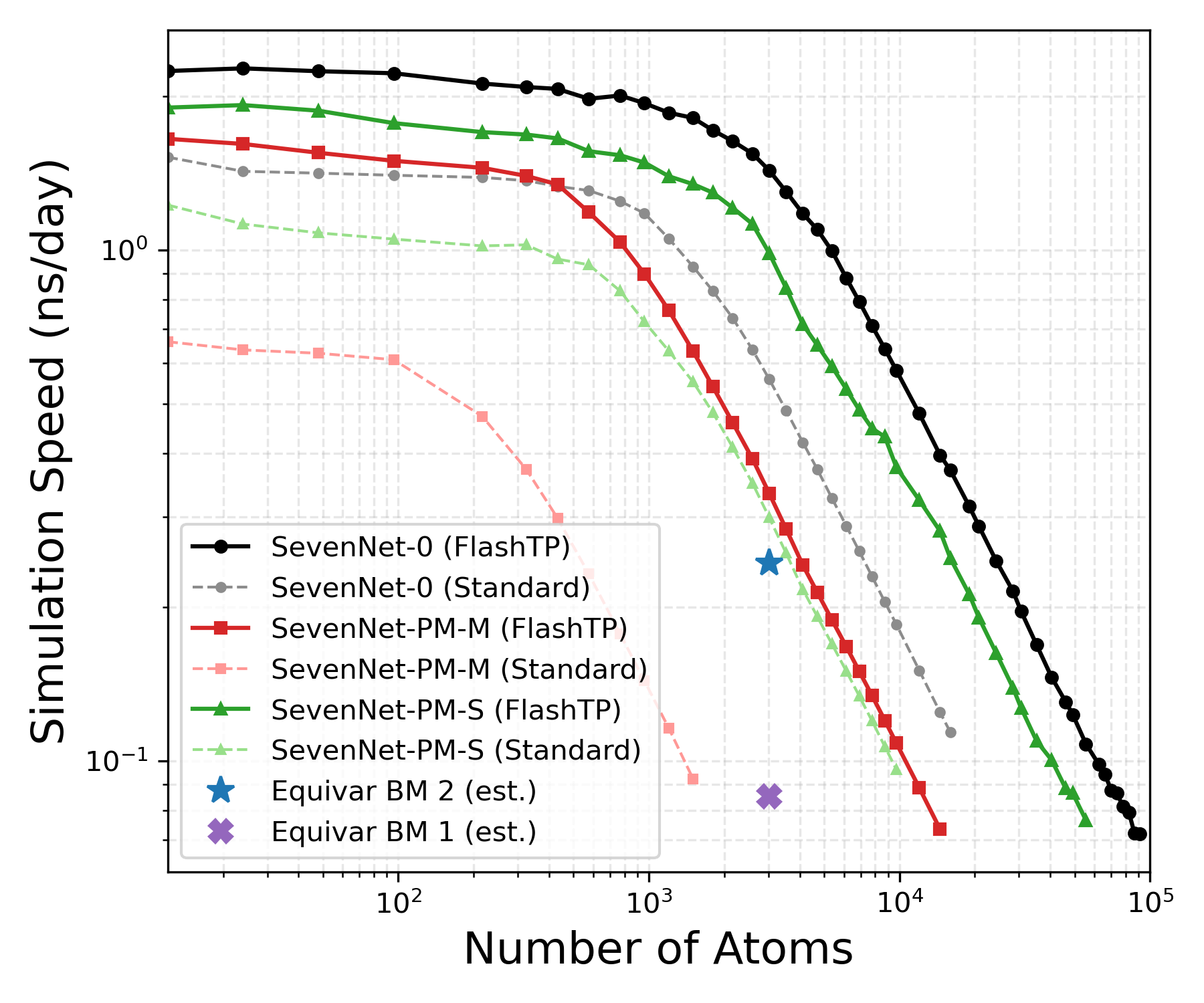}
                \caption{Simulation speed with respect to the number of atoms, tested on an NVIDIA GH200 GPU. The time was averaged over 100 MD steps using ASE. Estimated reference points for \texttt{equivar} BM 1 and BM 2 models at 3,024 atoms are included for comparison, assuming that the simulation time is entirely driven by the time for the BEC predictions.} \label{fig:Speed_vs_Atoms_GH200}
            \end{figure}
    
            \autoref{fig:Speed_vs_Atoms_GH200} shows the simulation speed achieved using SevenNet-Polar models. SevenNet-0 is shown as a reference. The impact of \texttt{FlashTP} is significant, as it leads to speedup in the range of $\times$3 to $\times$5, while the reduction in memory footprint enables system size to be increased by $\times$5. Because our implementation enforces tensor output for the last layer, and adds complexity to enable the prediction of the BEC tensor, the simulation speed is lower and the maximum system size is smaller than for SevenNet-0. A speed of 1 ns/day can be obtained with up to around 800 atoms for SevenNet-PM-M, and around 3,000 atoms for SevenNet-PM-S. This is more than 11 times faster than the estimated speed for \texttt{equivar}\cite{Kutana_2025}, whose inference speed for only BEC (no energy/force/stress nor time integration) was measured as 0.340 ms/atom for BM1 and 0.117 ms/atom for BM2, leading to an estimation of 1 s/step for ~3000 atoms.  For a 3,024-atom system, we achieve an inference time of 0.086 ms/atom for SevenNet-PM-M and 0.029 ms/atom for SevenNet-PM-S. Therefore, despite predicting not only BEC but also energy, forces and stress, our inference time per atom remained consistently faster than the fastest inference time (0.117 ms/atom) reported for \texttt{equivar}\cite{Kutana_2025}, while achieving better accuracy. In terms of maximum system sizes,, when running on a single GH200 GPU (96 GB VRAM), without \texttt{FlashTP}, the SevenNet-PM-M and SevenNet-PM-S models are limited to system sizes of 1,500 and 9,720 atoms, respectively. The use of \texttt{FlashTP} allows pushing the limits to 14,520 and 55,488 atoms, respectively, for this memory capacity. More details can be found in \cref{tab:bec_overhead}, \cref{sec:bec_overhead} of the Supplementary Information. 
            
\begin{figure}[htbp]
                \centering
                \includegraphics[width=\linewidth]{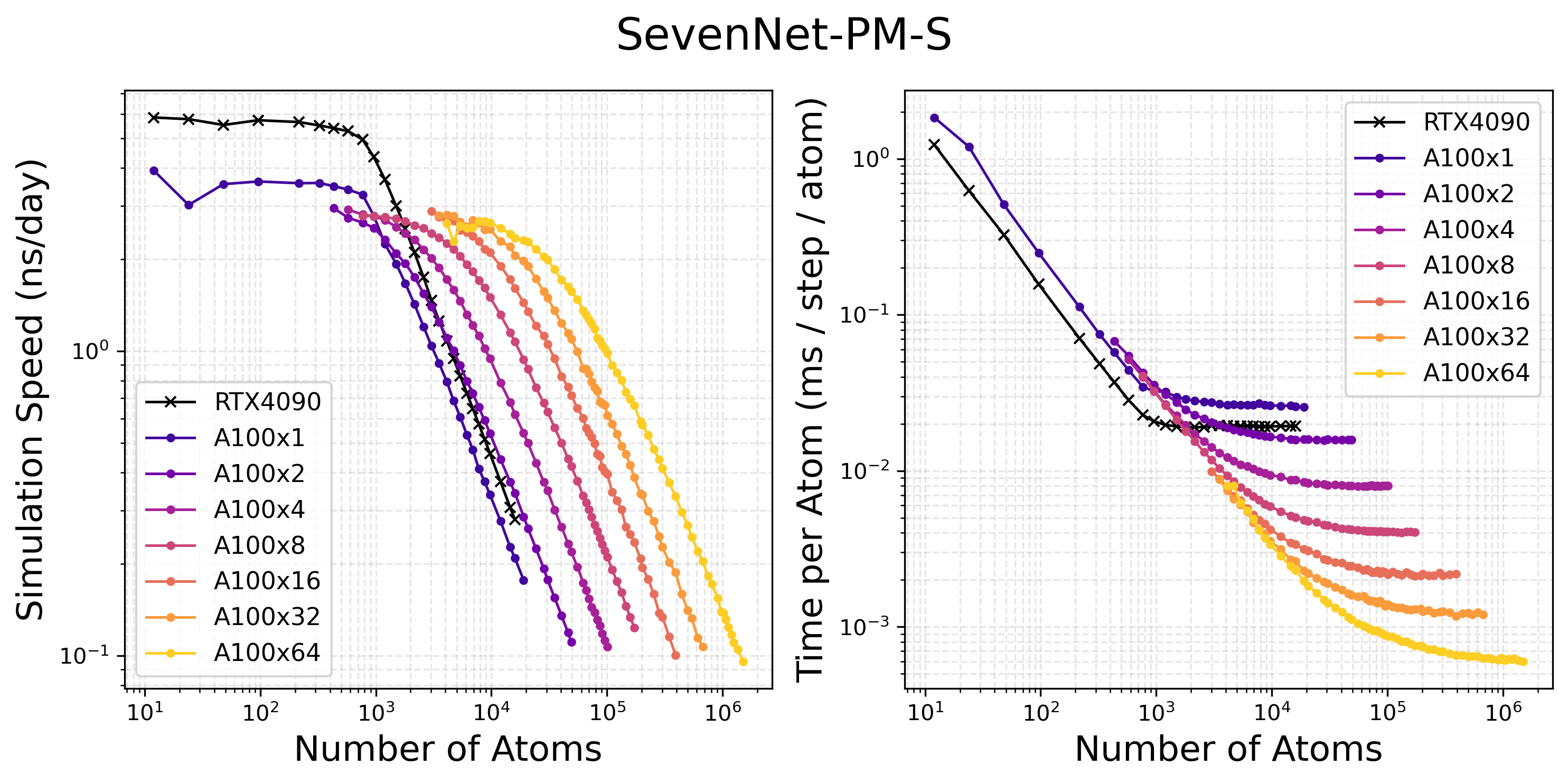}
                \caption{Evolution of the simulation speed (left) and inference time per atom (right) with respect to the number of GPUs in use. The black line with cross markers shows the reference consumer-grade NVIDIA GeForce RTX 4090.}
                \label{fig:GPU_scaling_SevenNet}
            \end{figure}

            The implementation of a \texttt{LAMMPS} interface enables the use of multiple GPUs, therefore unlocking faster and larger simulations, as can be seen in \autoref{fig:GPU_scaling_SevenNet}. The fastest speed obtained on a single consumer GPU (NVIDIA GeForce RTX 4090, 24 GB VRAM) was 0.098 ms/atom on SevenNet-PM-M (1,500 atoms) and 0.019 ms/atom on SevenNet-PM-S (2,160 atoms), while on the server-grade GPUs, we reached 0.0028 ms/atom on SevenNet-PM-M (64$\times$ A100, 40GB VRAM each, 345,960 atoms) and 0.0006 ms/atom on SevenNet-PM-S (64$\times$ A100, 1,500,000 atoms). In both cases, with 64$\times$ A100, the simulation speed remained at 1.0-1.1 step/s. The maximum size for each GPU configuration can be found in Section S11 of the Supplementary Information. Our methodology, by enabling faster and larger simulations including BEC, makes these calculations more accessible, as regular consumer grade GPUs can now perform simulations of 1 ns/day with several thousands of atoms, while on high-performance computing infrastructure, one can maintain this simulation speed up to around 100,000 atoms.

    \subsection{Applications}      
    
        \subsubsection{Nudged elastic band (NEB) calculations under electric field in tetragonal-\texorpdfstring{\ce{ZrO2}}{ZrO2}}
        
            \begin{figure}[htbp]
                \centering
                \includegraphics[width=0.3\linewidth]{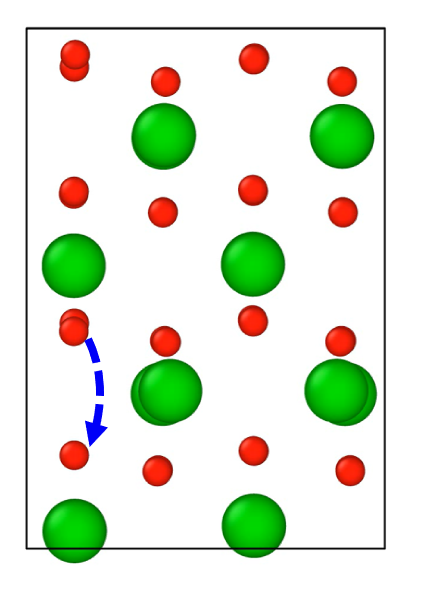}
                \includegraphics[width=0.65\linewidth]{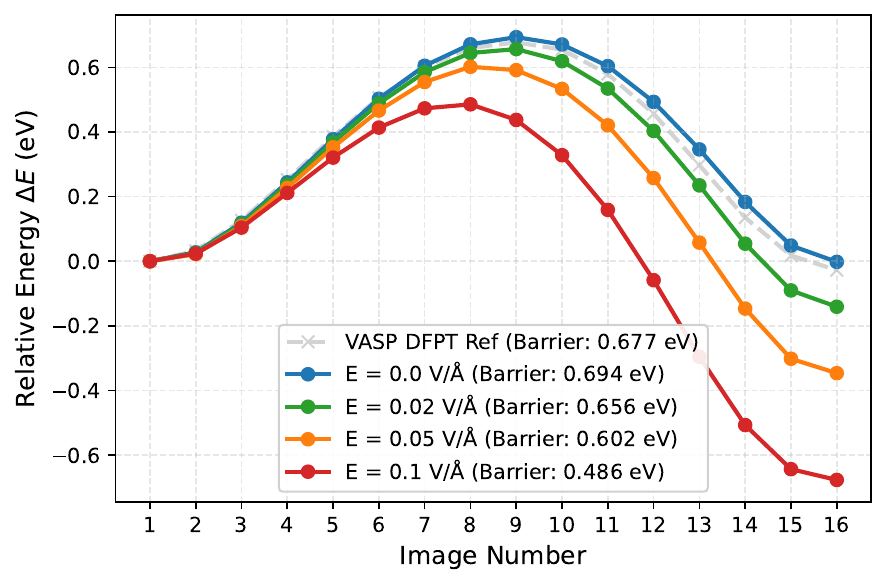}
                \caption{Nudged elastic band (NEB) calculations with SevenNet-PM-M for O diffusion to a vacancy site, with an electric field up to 0.1 V/\AA~along z direction. For reference, the DFT energy profile is shown in gray, illustrating the agreement in energy barrier. It should be noted that for a given image number, the structure differs between DFT and our model.}
                \label{fig:neb_004_field_sweep_profile}
            \end{figure}
        
            A direct application of SevenNet-PM-M is nudged elastic band (NEB) calculations for O migration in \ce{ZrO2}, similarly to the works of Shimizu~\cite{Shimizu_2023} and Lu \cite{Lu_2026}, but now using a single model with higher accuracy. We perform climbing image NEB calculations using ASE~\cite{Henkelman_2000,Henkelman_2000_2}.  \autoref{fig:neb_004_field_sweep_profile} shows NEB calculations for O diffusion to a vacancy site. The activation energy $E_A$ calculated by our model matches the DFT results (0.677 eV) within 0.02 eV with a calculated barrier energy of 0.694 eV. Consistent with previous results by Lu et al. \cite{Lu_2026}, applying an electric field along the z direction lowers the energy barrier. For each DFT NEB image, we also confirmed that the atom-resolved BEC prediction errors remain small. Details can be found in \cref{sec:neb_supplementary} of the Supplementary Information.

        \subsubsection{Evaluation on \texorpdfstring{$\Sigma 5(310)/[001]$}{Sigma 5(310)/[001]} grain boundary in cubic-\texorpdfstring{\ce{ZrO2}}{ZrO2}}
        
            \begin{figure}[htbp]
                \centering
                \includegraphics[width=0.45\linewidth]{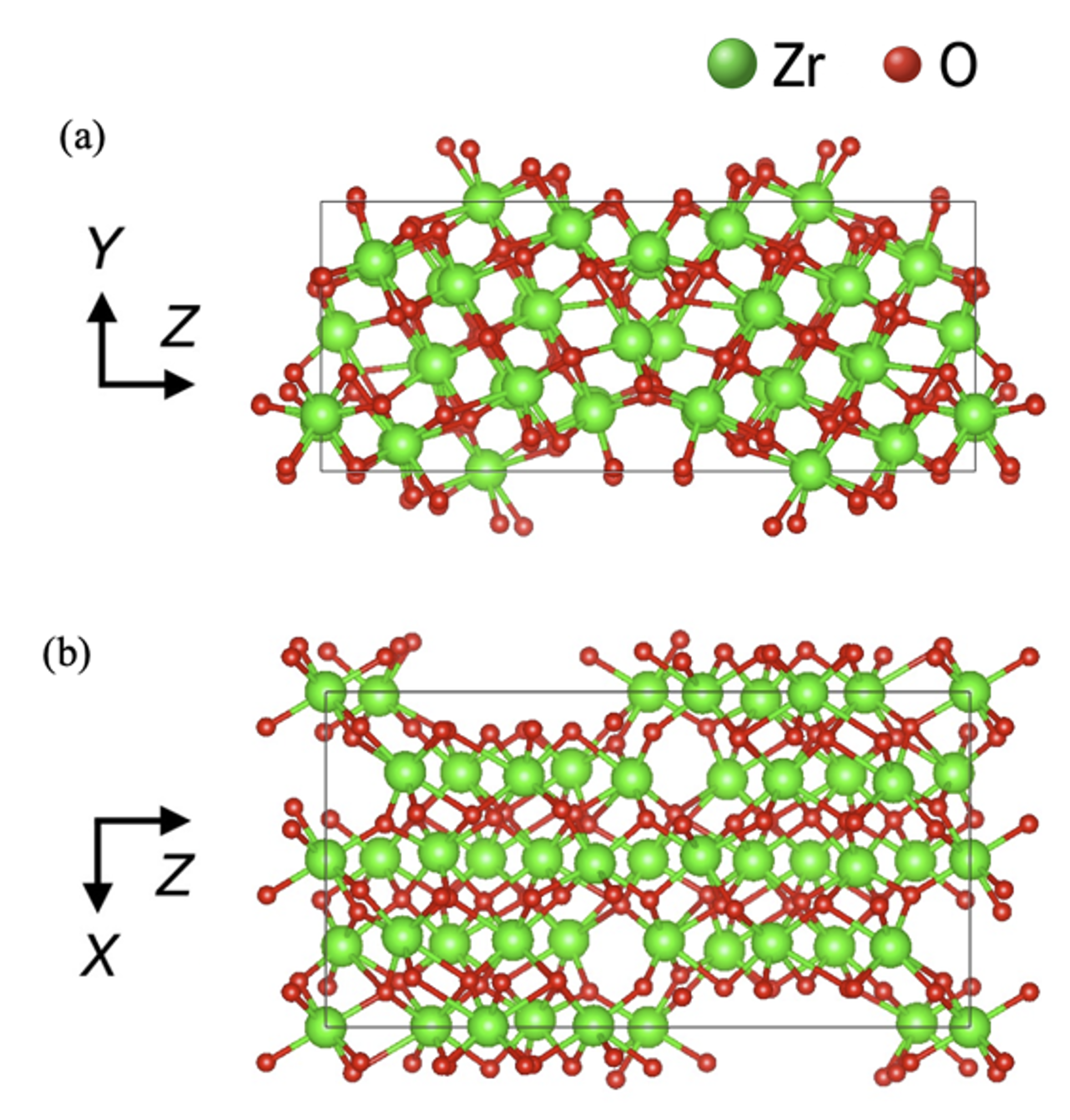}
                \includegraphics[width=0.45\linewidth]{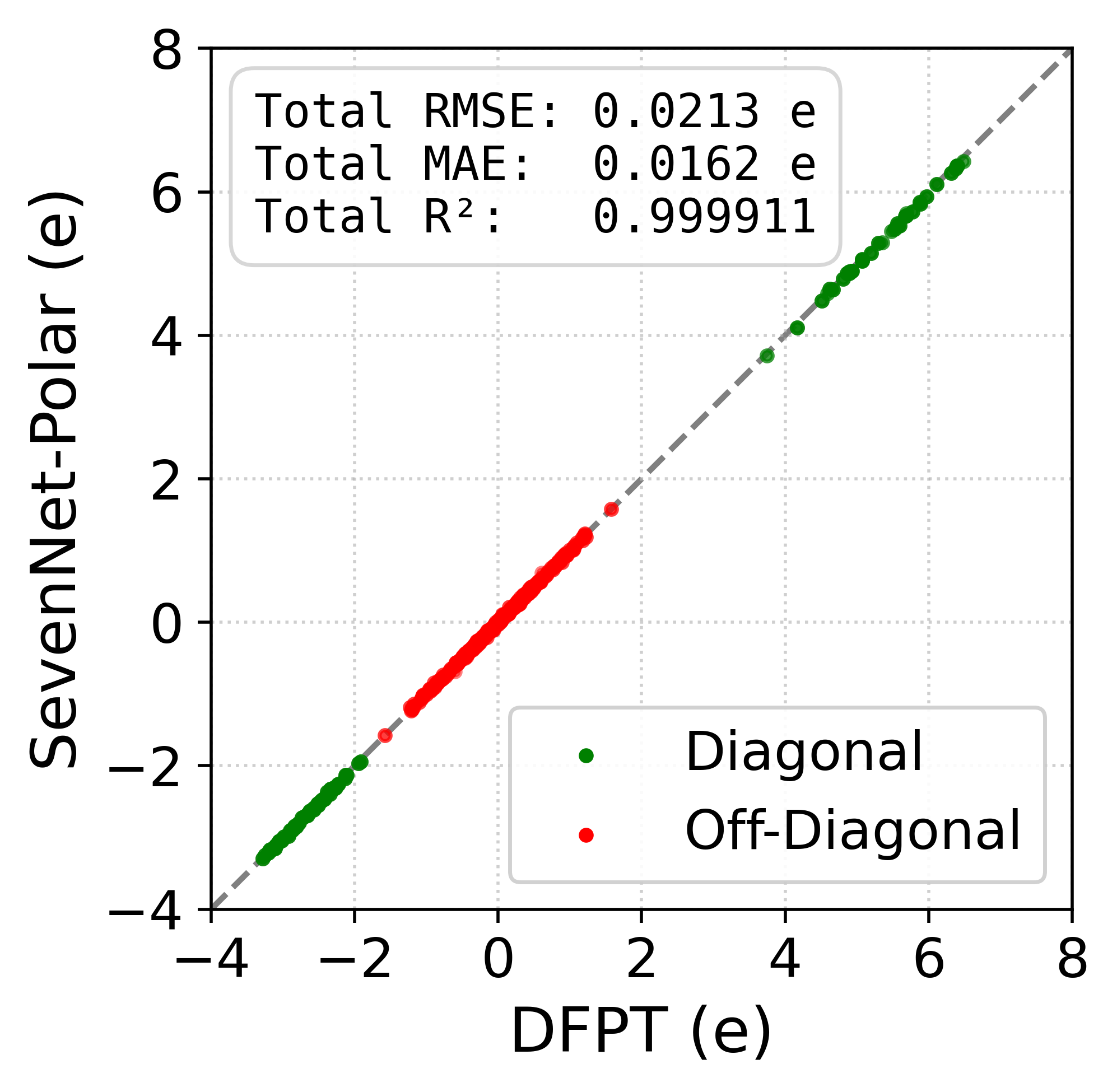}
                \caption{Prediction of the BEC tensor by SevenNet-PM-M on a $\Sigma 5(310)/[001]$ grain boundary in cubic-\ce{ZrO2} studied by Arai et al.\cite{Arai_2025}}
                \label{fig:gb1_bec_summary}
            \end{figure}
        
            SevenNet-PM-M was used to predict the BEC tensor on a $\Sigma 5(310)/[001]$ grain boundary in cubic-\ce{ZrO2} in a model presented by Arai et al.\cite{Arai_2025}. As can be seen in \autoref{fig:gb1_bec_summary}, despite the strong distortion in the vicinity of the grain boundary, the RMSE remains low, at 0.021 e, while the parity plot indicates reasonable transferability to this unseen extended defect, with no points far from the $y=x$ line. Detailed comparison between BEC calculated by DFPT and by the SevenNet-PM-M model is presented in \cref{sec:gb1_supplementary} of the Supplementary Information.

        \subsubsection{SevenNet-PS-M on water dimers, liquid water, \texorpdfstring{\ce{MAPbI3}}{MAPbI3}, liquid \texorpdfstring{\ce{NaCl}}{NaCl}, and \texorpdfstring{\ce{ZrO2}}{ZrO2} data set (Schmiedmayer et al. 2026 \texorpdfstring{\cite{Schmiedmayer_2026, Schmiedmayer_2026_data}}{})}
            We trained a SevenNet-PS-M model (BEC prediction only, $l_{\text{max}} = 3$) on the data set built by Schmiedmayer et al. \cite{Schmiedmayer_2026,Schmiedmayer_2026_data}, which includes solid, liquid, and dimer structures. The results, shown in  \autoref{tab:schmiedmayer_bec_rmse}, reveal that our model can reach lower RMSE values with a reduction of around 50\%--70\% compared to both the monopole-based models and MACE \cite{Batatia_2022}, although it is more complex and computationally intensive than the scalar monopole model proposed by Schmiedmayer et al.
        
        \begin{table*}[htbp]
            \centering
            \caption{Root Mean Square Error (RMSE) of Born effective charges predicted by SevenNet-Polar on the reference test set, alongside reference monopole and MACE models. Errors are reported in m$|e|$.}
            \label{tab:schmiedmayer_bec_rmse}
            \begin{tabular}{@{}lcccc@{}}
                \toprule
                & \multicolumn{2}{c}{\textbf{This Work}} & \multicolumn{2}{c}{\textbf{Reference Models \cite{Schmiedmayer_2026}}} \\
                \cmidrule(lr){2-3} \cmidrule(lr){4-5}
                \textbf{System} & \textbf{SevenNet-PS-M} & \textbf{SevenNet-PS-L} & \makecell{\textbf{Monopole + dipole} \\ \textbf{($q+p$)}} & \textbf{MACE} \\
                \midrule
                H$_2$O (liquid \& dimer) & 6.9  & 6.3 & $\sim$17  & $\sim$18 \\
                MAPbI$_3$ (multi-phase) & 9.5  & 8.2 & $\sim$35  & $\sim$35 \\
                NaCl (liquid)            & 10.7 & 9.2 & $\sim$20  & $\sim$19 \\
                ZrO$_2$ (solid)          & 41.2 & 37.4 & $\sim$110 & $\sim$95 \\
                \midrule
                \textbf{Global Evaluation} & 18.6 & 16.7 & --        & --        \\
                \bottomrule
            \end{tabular}
        \end{table*}
        
        \subsubsection{SevenNet-PM-L on \texorpdfstring{\ce{SiO2}}{SiO2} and \texorpdfstring{\ce{BaTiO3}}{BaTiO3} data sets (Falletta et al. 2025 \texorpdfstring{\cite{Falletta_2025}}{})}
        
            \begin{figure}[htbp]
                \centering
                \includegraphics[width=\linewidth]{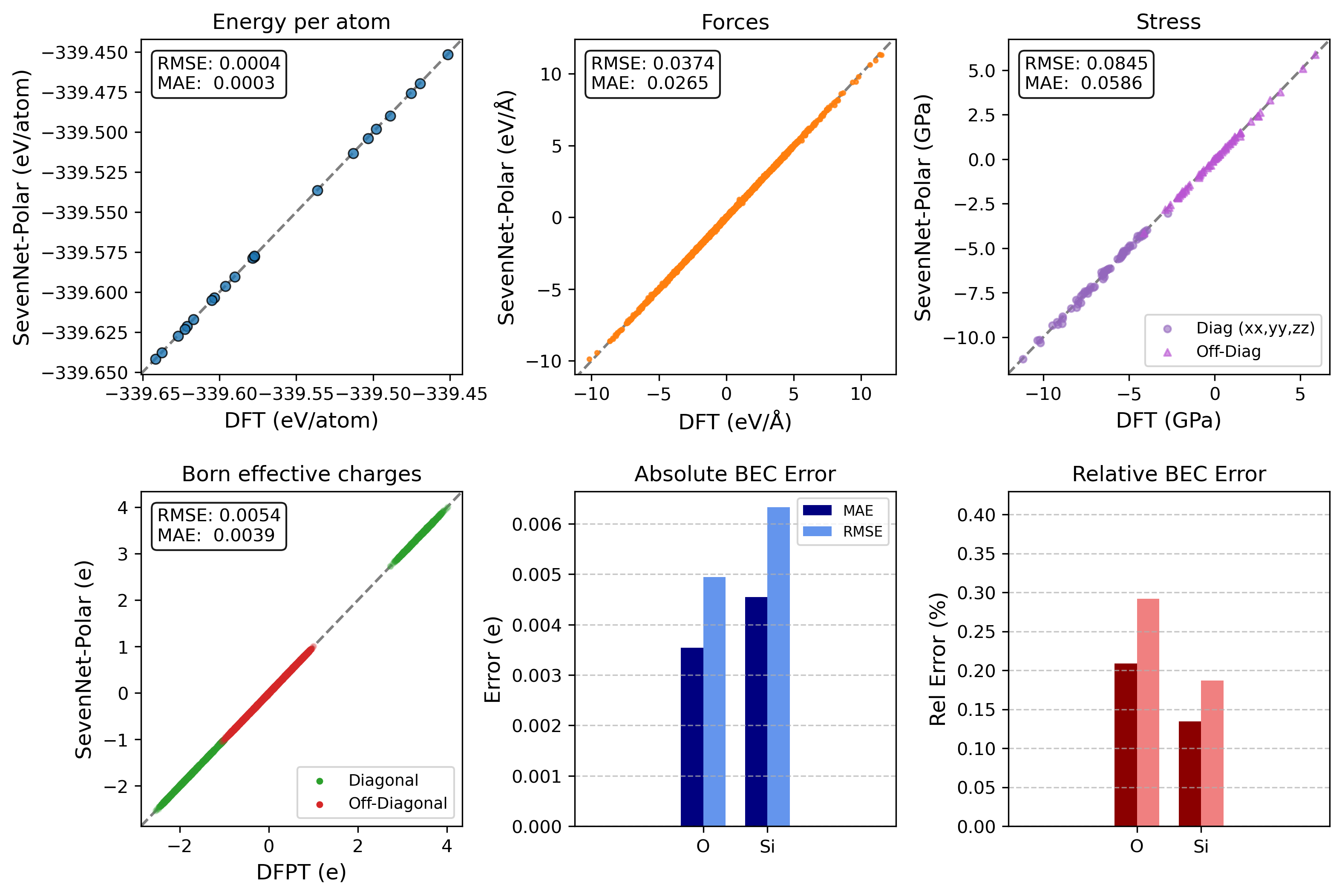}
                \caption{Parity plot for SevenNet-PM-L, evaluated on the test set of a \ce{SiO2} data set \cite{Falletta_2025}}
                \label{fig:SiO2_allegropol_test_Academic}
            \end{figure}

\begin{figure}[htbp]
                \centering
                \includegraphics[width=\linewidth]{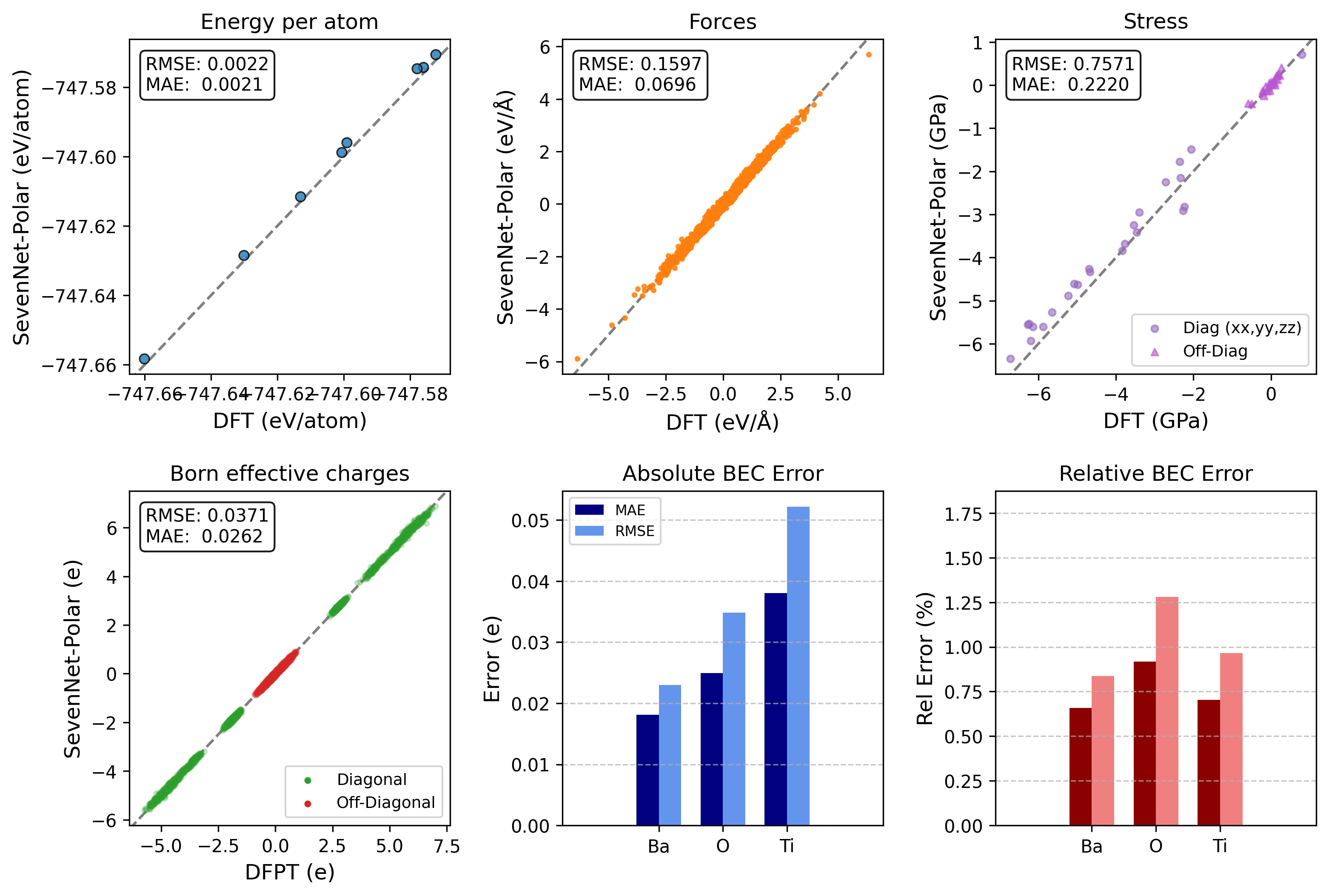}
                \caption{\ce{BaTiO3} data set \cite{Falletta_2025}, evaluated on the test set.}
                \label{fig:multitask_batio3_test_SevenNet}
            \end{figure}

            We trained a SevenNet-PM-L model on the \ce{SiO2} data set generated by Falletta et al. \cite{Falletta_2025} which contains 201 structures. The data was split 160/20/21 into train/validation/test sets. For the training (test) set, as shown in \autoref{fig:SiO2_allegropol_test_Academic}, we obtained a mean absolute error (MAE) of 0.4 (0.3) meV/atom on energy, 26.9 (26.5) meV/\AA~ on forces, 0.063 (0.059) GPa on stress and 0.0032 (0.0039) on the BEC tensor, which is equivalent to the accuracy achieved by Falletta et al. \cite{Falletta_2025} with the \texttt{allegro-pol} model. 
            On the \ce{BaTiO3} training (resp. test) set generated by Falletta et al. \cite{Falletta_2025}, which contains 73 structures (58/7/8 split), we obtained a mean absolute error (MAE) of  0.6 (0.6) meV/atom on energy, 71.9 (72.5) meV/\AA~ on forces, 0.148 (0.130) GPa on stress and 0.0135 (0.0176) on the BEC tensor, as shown in \autoref{fig:multitask_batio3_test_SevenNet}. These values are larger than those reported by Falletta et al. with the \texttt{allegro-pol} model. Nevertheless, a study of the scaling laws presented in \cref{sec:scaling_law_sio2_batio3} of the Supplementary Information reveals that better accuracy could be reached if this data set is expanded with more atomic configurations. This shows that our model can also achieve fairly good accuracy in data-starved regime, as the smaller set has less than 100 structures. 
            
\section{Discussion}

    We enabled accurate and fast prediction of the Born effective charge tensor with SevenNet-Polar, based on the SevenNet architecture, marking an extension beyond per-atom scalar and vector predictions. Our largest SevenNet-PM-L model achieves excellent accuracy across all quantities for high-temperature and defect-containing \ce{ZrO2} and \ce{Li3PO4} configurations, with typical root mean squared errors (RMSE) of less than 1.0~meV/atom for energy, 11.9~meV/\AA~for forces, 0.05~GPa for stress, and below 0.005~$e$ for the BEC tensor (for both diagonal and off-diagonal terms). The scaling analysis shows that different quantities follow different power laws ($RMSE\propto N^{-\alpha}$), with the exponent $\alpha$ being approximately 0.73 for energy, 0.53 for forces, 0.72 for the stress tensor, 0.39 for the diagonal terms of the BEC tensor, and 0.32 for its off-diagonal terms. This demonstrates not only that the BEC tensor is inherently more difficult to predict, but also that the off-diagonal terms are the most challenging to train. A hyperparameter study confirmed that this higher accuracy originates from a combination of higher angular momentum for the spherical harmonics ($l_{\text{max}}$) and an extended receptive field (6~\AA~$\times$ 5 layers). However, increasing the number of layers $L$ and the maximum order for the spherical harmonics ($_{max}=4$) leads to a dramatic increase of the calculation time. Therefore, more efficient models that combine accuracy and speed are proposed with, SevenNet-PS-M and SevenNet-PM-M  (2 to 3 million parameters), though we also propose lighter architectures with reduced complexity ($l_{\text{max}}=2$), SevenNet-PS-S and SevenNet-PM-S (400,000 to 600,000 parameters). 

    The SevenNet-PM multitask model family removes the need to combine separate NNP and BEC models as done in previous works~\cite{Shimizu_2023,Kutana_2025,Lu_2026}, though compared to \texttt{equivar}, a larger cutoff beyond 3~\AA~is needed in order to achieve good accuracy on the forces and stress tensor. SevenNet-PM-M demonstrated robust predictive capabilities when tested on NEB calculations of oxygen migration in defect-laden tetragonal \ce{ZrO2} and for BEC predictions of a $\Sigma 5 (310)$ grain boundary in cubic-\ce{ZrO2}. This model can also be used for molecular dynamics simulations of systems under an applied electric field. 
    
    Our implementation is compatible with \texttt{FlashTP}, and implementation of \texttt{ASE} calculators and \texttt{LAMMPS} interface enable GPU-accelerated molecular dynamics simulations with more than 15,000 atoms on a consumer-grade NVIDIA GeForce RTX 4090, and more than 100,000 atoms on a single server-grade A100 GPU, significant improvement over the baseline limits of 2,592 and 14,520 atoms without \texttt{FlashTP}, respectively. For computational context, while prior work using \texttt{equivar} reported speeds exceeding 1~s/step for a 3,000-atom system when predicting only the BEC tensor, our approach computes full molecular dynamics steps at a rate of 3~steps/s with SevenNet-PM-M, and 11~steps/s with SevenNet-PM-S in single-GPU settings. Multi-GPU simulations are enabled with \texttt{LAMMPS}, allowing simulation speeds below 0.001 ms/atom/step and system sizes that can reach 1,500,000 atoms on 64 A100 GPUs, the main limiting factor being memory.
    
    A possible application is to consider variable electric fields with ferroelectric materials such as \ce{BaTiO3}, which exhibit an hysteresis loop, as calculated by Chen and Mizoguchi ~\cite{Chen_2026} using machine learning potentials. The present architecture would enable simulations of larger models, which are necessary to better capture the formation of ferroelectric domain walls. In the future, an additional extension of the SevenNet architecture to predict the macroscopic dielectric tensor $\epsilon_{\infty}$ would enable phonon calculations including non-analytical term corrections (NAC) via packages such as \texttt{phonopy}~\cite{Togo_2015,Togo_2023}. 
    
    The present work broadens access to the use of charge-aware neural networks for molecular dynamics simulations under an electric field, making it accessible on mainstream GPUs for systems of thousands of atoms at practical speeds ($\geq 1$~ns/day).  An extension to a larger number of atomic species towards a universal BEC model could also be built, if strategies such as improved architectures, active learning, or other can mitigate the unfavorable BEC scaling exponents.

\section{Methods}
  
    \subsection{Scalable EquiVariance-Enabled Neural Network (SevenNet)}

        The Scalable Parallel Algorithm for Graph Neural Network (SevenNet\cite{Park_2024}) is a promising architecture based on NequIP \cite{Batzner_2022}, which has demonstrated excellent performance as a uMLIP. The recent SevenNet-Omni \cite{Kim_2026} model ranks among the top-performing models on the Matbench Discovery benchmark \cite{Riebesell_2025}. In this architecture, irreducible representations (irreps) of the $O(3)$ group are used in the network layers via Clebsch-Gordan tensor products \cite{Varshalovich_1988}. While the prediction of scalar or vector quantities is well-established, and global tensor quantities such as the stress tensor can also be predicted, this work expands the predictive complexity of SevenNet by introducing the prediction of a per-atom rank-2 tensor: the Born effective charge tensor.
        
    \subsection{Born effective charges (BEC)}
    
         The Born effective charge (BEC) tensor, $Z^*_{i,\alpha\beta}$, describes the change in macroscopic polarization $\mathbf{P}$ induced by the displacement of atom $i$ in direction $\beta$ \cite{Born_1954}:
        \begin{equation}
            Z^*_{i,\alpha\beta} = \Omega \frac{\partial P_\alpha}{\partial u_{i,\beta}}
        \end{equation}
        where $\Omega$ is the cell volume. Here, we express the BEC tensor in units of elementary charge $e$. The BEC tensor is a typical quantity obtained from density functional perturbation theory (DFPT) \cite{Gonze_1997}. In this work, we extend the SevenNet framework to predict Born effective charges, exploiting the strict equivariance embedded in its architecture. We demonstrate how to encode, propagate, and predict these $3 \times 3$ tensors using irreducible representations (irreps) of $O(3)$, ensuring the physical correctness and geometric consistency of the predicted properties. Throughout our evaluation, we conduct separate error analyses for the diagonal and off-diagonal components of the BEC tensor.

    \subsection{FlashTP}

        While equivariant architectures like NequIP \cite{Batzner_2022}, Allegro \cite{Musaelian_2023}, MACE \cite{Batatia_2022}, and SevenNet \cite{Park_2024} provide state-of-the-art accuracy and preserve rigorous physical symmetries, they historically suffer from significant computational bottlenecks. The core operations, which are Clebsch-Gordan tensor products of $O(3)$ irreps, can account for up to 75\% of the total training time and introduce substantial memory overhead \cite{Lee_2025}. This bottleneck is further exacerbated in this work due to extending the predictive capacity to dense, per-atom rank-2 tensors like the $3 \times 3$ Born effective charge tensors. This  significantly increases the number of interacting feature paths and channels compared to scalar energy predictions.

        By natively integrating \texttt{FlashTP} into our framework, the peak memory footprint is significantly reduced, and we significantly accelerate both inference and training execution times. This computational efficiency is critical for scaling the prediction of complex tensorial properties like BECs to larger atomistic systems. It ensures we can maintain the exact geometric consistency and strict $E(3)$-equivariance of the architecture without incurring the prohibitive computational costs typically associated with high-degree irreducible representations. For the BEC predictions, \texttt{FlashTP} reduces the computation overhead from more than 100\% with standard \texttt{e3nn} to only 30 \% or less (details in \cref{sec:bec_overhead} of the Supplementary Information). This ultimately enabled single-GPU molecular dynamics simulations including BEC predictions for up to 14,520 atoms on an RTX 4090 and 55,488 atoms on a GH200 using SevenNet-PM-S. A large speedup was observed at equal number of atoms, as summarized by \autoref{tab:flashtp_impact}. Benchmarks using \texttt{LAMMPS} (multi-GPU) and \texttt{FlashTP} with a large selection of GPUs are shown in \cref{sec:detailed_benchmark_gpu} of the Supplementary Information.
        
        \begin{table*}[t]
            \caption{\label{tab:flashtp_impact} Impact of \texttt{FlashTP} optimization on maximum system size and on the simulation speed with ASE. 10 warming-up MD time steps were performed, followed by 100 MD time steps for which timing was done. Performance is compared across a consumer GPU (NVIDIA GeForce RTX 4090, 24 GB) and an enterprise-grade superchip (NVIDIA GH200, 96 GB) for four architectures: SevenNet-0, SevenNet-Omni-i12, SevenNet-PM-S, and SevenNet-PM-M. Max N denotes the largest atomic system before out-of-memory (OOM) occurs. Simulation speeds are compared for a simulation size of 2,592 atoms, as well as the speed-up when data is available.}
            \footnotesize 
            \begin{tabular}{@{}lll cc@{}}
            \toprule
            \textbf{Model} & \textbf{Hardware} & \textbf{\texttt{FlashTP}} & \textbf{Max N} & \textbf{Simulation speed with 2,592 atoms} \\
            & & & (atoms) & (ns/day) \\
            \midrule
            \multirow{4}{*}{\textbf{SevenNet-0}} & \textbf{RTX 4090} & No & 4,116 & 0.10 \\
            & & Yes & \textbf{20,736} (5.0$\times$) & \textbf{1.84} (18.4$\times$) \\
            \cmidrule{2-5}
            & \textbf{GH200} & No & 15,972 & 0.21 \\
            & & Yes & \textbf{91,200} (5.7$\times$) & \textbf{1.55} (7.4$\times$) \\
            \midrule
            \multirow{4}{*}{\textbf{SevenNet-Omni-i12}} & \textbf{RTX 4090} & No & 324 & N/A \\
            & & Yes & \textbf{2,592} (8.0$\times$) & \textbf{0.20} (N/A) \\
            \cmidrule{2-5}
            & \textbf{GH200} & No & 1,500 & N/A \\
            & & Yes & \textbf{9,720} (6.5$\times$) & \textbf{0.26} (N/A) \\            
            \midrule
            \multirow{4}{*}{\textbf{SevenNet-PM-S}} & \textbf{RTX 4090} & No & 2,592 & 0.15 \\
            & & Yes & \textbf{14,520} (5.6$\times$) & \textbf{1.15} (7.6$\times$) \\
            \cmidrule{2-5}
            & \textbf{GH200} & No & 9,720 & 0.35 \\
            & & Yes & \textbf{55,488} (5.7$\times$) & \textbf{1.12} (3.2$\times$) \\
            \midrule
            \multirow{4}{*}{\textbf{SevenNet-PM-M }} & \textbf{RTX 4090} & No & 432 & N/A \\
            & & Yes & \textbf{3,528} (8.2$\times$) & \textbf{0.31} (N/A) \\
            \cmidrule{2-5}
            & \textbf{GH200} & No & 1,500 & N/A \\
            & & Yes & \textbf{14,520} (9.7$\times$) & \textbf{0.39} (N/A) \\
            \botrule
            \end{tabular}
        \end{table*}
            
    \subsection{Implementation of the Born effective charge tensor in SevenNet}
  
        Because a standard Cartesian rank-2 tensor transforms as a reducible representation under three-dimensional rotations, predicting the $3 \times 3$ Born effective charge (BEC) matrix directly without enforcing the proper transformation law would not guarantee rotational equivariance~\cite{Thomas_2018}. To preserve $O(3)$ equivariance within the SevenNet architecture, the target BEC tensor must be decomposed into a direct sum of irreps~\cite{Geiger_2022}. Mathematically, a general, non-symmetric rank-2 tensor decomposes into three irreducible equivariant subspaces:
        \begin{itemize}
            \item A scalar trace (isotropic component, $L=0$): $\mathbf{1 \times 0e}$
            \item An antisymmetric vector (pseudo-vector component, $L=1$): $\mathbf{1 \times 1e}$
            \item A symmetric traceless tensor ($L=2$): $\mathbf{1 \times 2e}$
        \end{itemize}
        Using the \texttt{e3nn} framework~\cite{Geiger_2022}, we define the target readout representation for the BEC tensor as $\Gamma_{\text{BEC}} = \mathbf{1 \times 0e + 1 \times 1e + 1 \times 2e}$, comprising 9 degrees of freedom. During training, the ground-truth density functional theory (DFT) Cartesian tensors are projected onto this spherical harmonic basis via Clebsch-Gordan coefficients, preserving the exact $L_2$ norm and ensuring geometrically consistent multitask optimization~\cite{Falletta_2025}. Naturally, fully reconstructing this decomposition requires setting the maximum angular resolution parameter, $\mathbf{l_{max}}$, to 2 or higher. We also evaluate models with $\mathbf{l_{max}}=0,1$ to demonstrate the impact of truncating these higher-order features.
        
       The computational cost of tensor operations is significantly mitigated by using \texttt{FlashTP} \cite{Lee_2025}, which accelerates training by a factor of 3-to-4, while simultaneously reducing memory usage. Detailed information about the implementation can be found in \cref{sec:ExtxyzBEC,sec:CodeChanges,sec:NewKeywords} of the Supplementary Information. A calculator for \texttt{ASE} and an interface for \texttt{LAMMPS} were written. Details can be found in \cref{sec:calculators_and_lammps} of the Supplementary Information.

    \subsection{Loss function}
    
        The loss function is defined as a weighted sum of individual property losses:
        \begin{equation}
        \label{eq:total_loss}
            \mathcal{L}_{\text{total}} = w_E \mathcal{L}_E + w_F \mathcal{L}_F + w_S \mathcal{L}_S + w_{\text{BEC}} \mathcal{L}_{\text{BEC}}
        \end{equation}
        
where $\mathcal{L}_E$, $\mathcal{L}_F$, $\mathcal{L}_S$, and $\mathcal{L}_{\text{BEC}}$ represent the mean squared errors (MSE) for energy, atomic forces, virial stress, and the Born effective charge tensors, respectively. The empirical loss weights were chosen to balance the vastly different magnitudes of the property gradients during backpropagation. Taking the energy scale as the implicit reference ($w_E = 1.0$), the force weight was set to $w_F = 0.1$, the stress weight to $w_S = 10^{-6}$, and the BEC weight to $w_{\text{BEC}} = 10.0$.

    \subsection{Model parameters and training conditions}

        To properly capture the complex angular dependency required for tensor predictions, our reference architecture employs a maximum angular momentum of $l_{\text{max}}=3$ and 64 feature channels per atom. The local atomic environments are defined by a cutoff distance of $r_c=6.0$~\AA, with geometric information propagated through $4$ equivariant message-passing layers.  An alternative, computationally lighter model ($l_{\text{max}} = 2$, $C=32$) is also evaluated for applications requiring faster inference.
        
        Model training was performed using in-house NVIDIA GeForce RTX 4080 SUPER, NVIDIA GeForce RTX 4090, NVIDIA RTX PRO 6000 Blackwell Max-Q, NVIDIA A800 40GB, and NVIDIA A100 80GB GPUs, as well as NVIDIA A100 40GB GPUs at the Wisteria/BDEC-01 supercomputer and NVIDIA GH200 GPUs at the Miyabi supercomputer, both operated by the Information Technology Center at The University of Tokyo. The present implementation is based on SevenNet \cite{Park_2024} 0.12.2.dev0 (as of March 29, 2026). Data sets are stored in the \texttt{extXYZ} file format, where all information about energy, forces, stress tensor and BEC tensors are written. SevenNet can natively read these files via \texttt{ASE}. For all models, the data sets were partitioned into a standard 80\% training, 10 \% validation, and 10\% test split. Unless otherwise stated, these represent the default training conditions throughout the following sections.

    \subsection{Computational details of the data sets}

        Three data sets are used in this work. The data was generated by density functional theory (DFT) calculations \cite{Hohenberg_1964,Kohn_1965} and density functional perturbation theory (DFPT) calculations \cite{Gonze_1997} using VASP~\cite{Kresse_1996}. The details are as follows.
        \begin{enumerate}
            \item \ce{ZrO2} data set by Lu et al. \cite{Lu_2026} of 10,103 structures obtained with the PBE exchange-correlation functional \cite{Perdew_1996}. Note that an older version of this data set was used by Kutana et al. \cite{Kutana_2025}, but has been rigorously refined to remove poorly converged calculations present in the original data set. Among these structures, 3,977 are pristine \ce{ZrO2}, while the remaining configurations incorporate point defects (such as oxygen vacancies) sampled across a high-temperature range of 1,000~K to 1,900~K. Structures under strains (from -2\% to +2\%) were also included. This inclusion is vital for training the model to predict the highly anisotropic, localized dielectric responses surrounding defect sites.
            \item \ce{Li3PO4} data set by Shimizu et al. \cite{Shimizu_2023} of 17,900 structures, obtained with the PBE exchange-correlation functional \cite{Perdew_1996}. This data set includes highly distorted configurations sampled from molecular dynamics trajectories at  high temperatures (up to 2,000~K), capturing the complex, fluctuating polarization landscape.
            \item Perovskite data set by Kutana et al. \cite{Kutana_2025} of 1,224 pristine structures, obtained with the PBESol exchange-correlation functional \cite{Perdew_2008}.            
        \end{enumerate}
    
        To ensure thermodynamic consistency regarding the exchange-correlation functional (PBE) for energy, force, and stress predictions, the SevenNet-PM family of multitask models are trained exclusively on the \ce{Li3PO4} and \ce{ZrO2} data sets. The SevenNet-PS family of specialized "BEC-only" models are trained on all three data sets.
\backmatter

\section*{Data availability}
The datasets and model checkpoints generated during this study are available in Zenodo at \url{https://doi.org/10.5281/zenodo.21322761}.

\section*{Code availability}
The source code for SevenNet-Polar is available at \url{https://github.com/AugustinLu/SevenNet-Polar}. The software package is available via Zenodo at \url{https://doi.org/10.5281/zenodo.21334102}.

\section*{Acknowledgments}
This work was supported by JSPS KAKENHI Grant Numbers 24K01284 and 26K08009. We also acknowledge the computational resources provided by the Information Technology Center at The University of Tokyo through the Wisteria/BDEC-01 and Miyabi systems. We also acknowledge the use of Gemini and Jules for assistance in manuscript drafting and the development of the software implementation used in this work.

\section*{Author contributions}
A.K.A.L. developed the SevenNet-Polar code, trained the models, conducted the atomistic simulations, and wrote the original manuscript. S.A. generated the grain boundary structures and provided the associated reference data. Y.P. contributed to the development of the multi-GPU LAMMPS integration critical for high-performance execution. S.H. contributed to the conceptualization of integrating Born effective charges into the architecture through extensive discussions. T.M. provided computational resources and engaged in detailed discussions regarding the methodology and applications. S.W. supervised the project, providing continuous guidance and technical discussions throughout the development cycle. All authors discussed the results, reviewed, and extensively revised the manuscript.

\section*{Competing interests}
The authors declare no competing interests.

\bibliography{sn-bibliography}
\clearpage
\setcounter{equation}{0}
\setcounter{figure}{0}
\setcounter{table}{0}
\setcounter{page}{1}
\setcounter{section}{0}
\renewcommand{\theequation}{S\arabic{equation}}
\renewcommand{\thefigure}{S\arabic{figure}}
\renewcommand{\thetable}{S\arabic{table}}

\begin{center}
  \Large\textbf{Supplementary Information for:}\\[0.5cm]
  \Large\textbf{SevenNet-Polar for MultiTask Prediction of Energy, Forces, Stress, and Born Effective Charges: Development and Application to \ce{ZrO2}, \ce{Li3PO4}, and Perovskites}\\[0.5cm]
  \normalsize Anh Khoa Augustin Lu, Shungo Arai, Yutack Park, Seungwu Han, Tsuyoshi Miyazaki, Satoshi Watanabe
\end{center}

\vspace{1cm}
\doublespacing
\section{Performance accuracy of \mbox{SevenNet-PS-S} and \mbox{SevenNet-PM-S}}\label{sec:sevennet_polar_small_models}

\begin{figure}[htbp]
    \centering
    \includegraphics[width=0.9\linewidth]{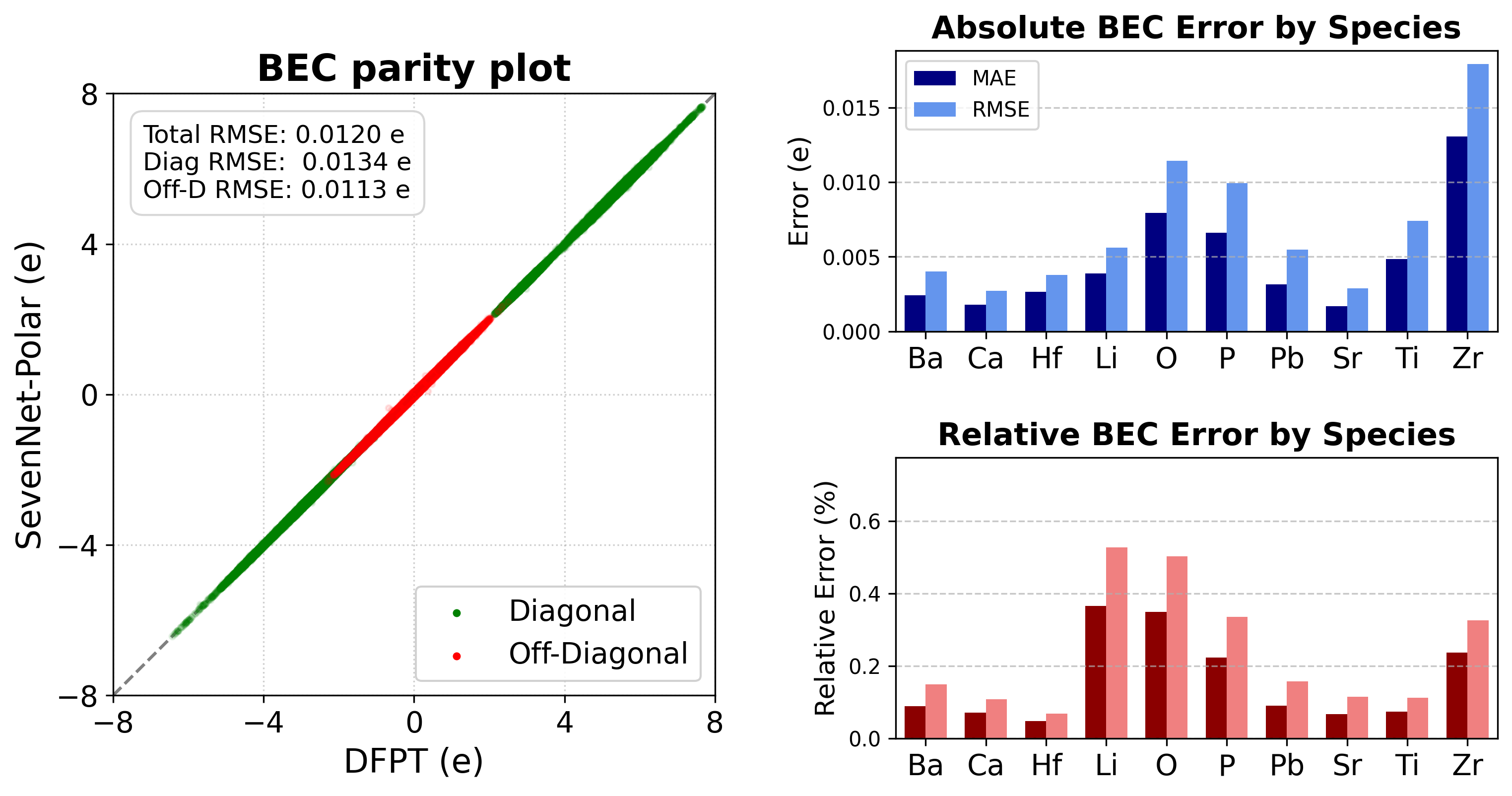}
    \caption{Performance accuracy of SevenNet-PS-S on the combined data set of \ce{ZrO2},  \ce{Li3PO4} and perovskites}
    \label{fig:specialized_light_on_triple_set_test}
\end{figure}

\begin{figure}[htbp]
    \centering
    \includegraphics[width=\linewidth]{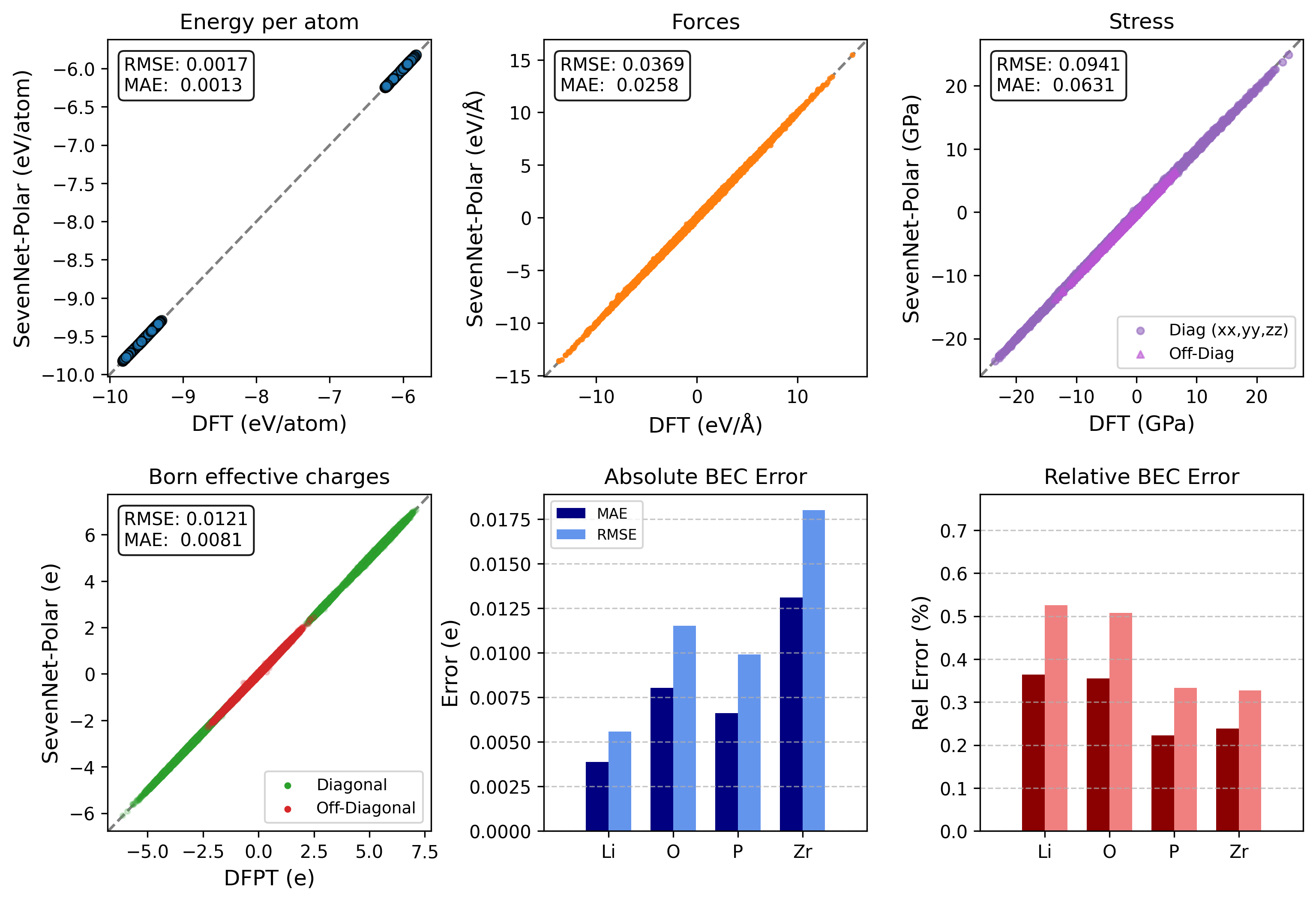}
    \caption{Performance accuracy of SevenNet-PM-S on the combined data set of \ce{ZrO2} and \ce{Li3PO4}}
    \label{fig:light_model_on_combined_Academic}
\end{figure}

The RMSE plots obtained with the light models are shown in \autoref{fig:specialized_light_on_triple_set_test} for SevenNet-PS-S and \autoref{fig:light_model_on_combined_Academic} for SevenNet-PM-S.

\clearpage

\section{Performance accuracy of \mbox{SevenNet-PS-M} on each data set}\label{sec:sevennet_ps_m_on_each_data_set}

\begin{figure}[htbp]
    \centering
    \includegraphics[width=0.9\linewidth]{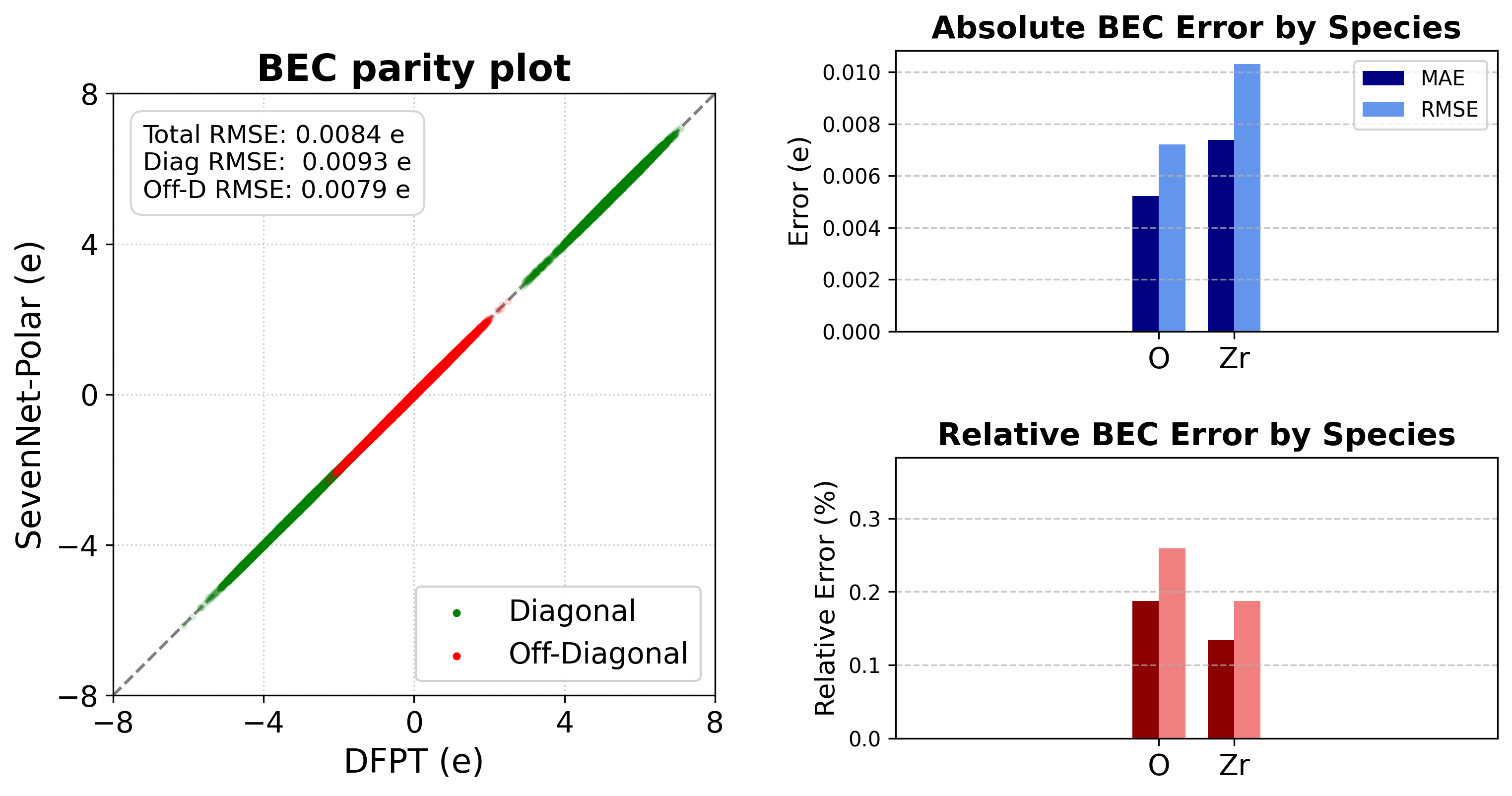}
    \caption{Evaluation of SevenNet-PS-M on the \ce{ZrO2} data set}
    \label{fig:specialized_on_zro2_test}
\end{figure}
\begin{figure}[htbp]
    \centering
    \includegraphics[width=0.9\linewidth]{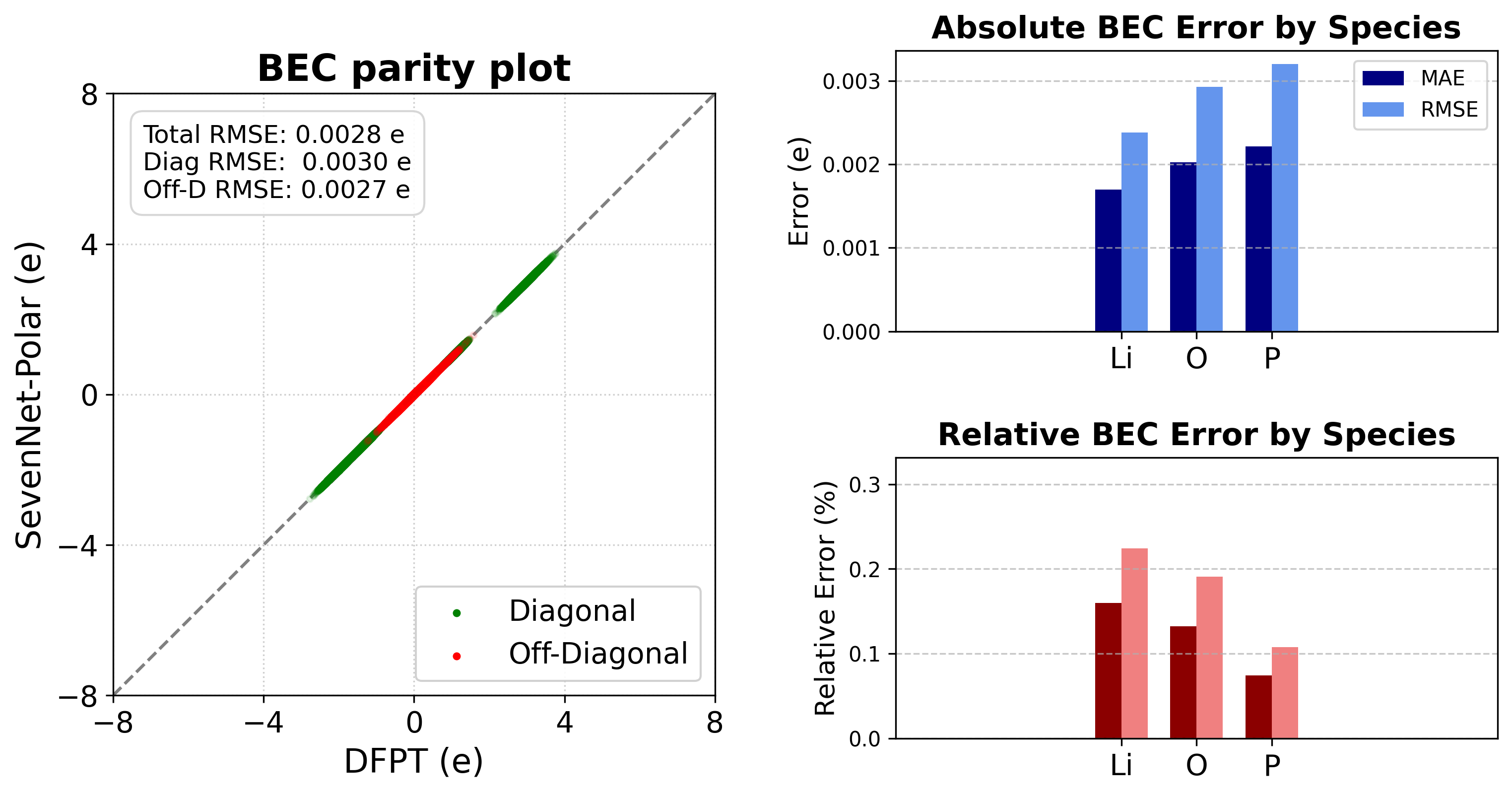}
    \caption{Evaluation of SevenNet-PS-M on the \ce{Li3PO4} data set}
    \label{fig:specialized_on_li3po4_test}
\end{figure}
\begin{figure}[htbp]
    \centering
    \includegraphics[width=0.9\linewidth]{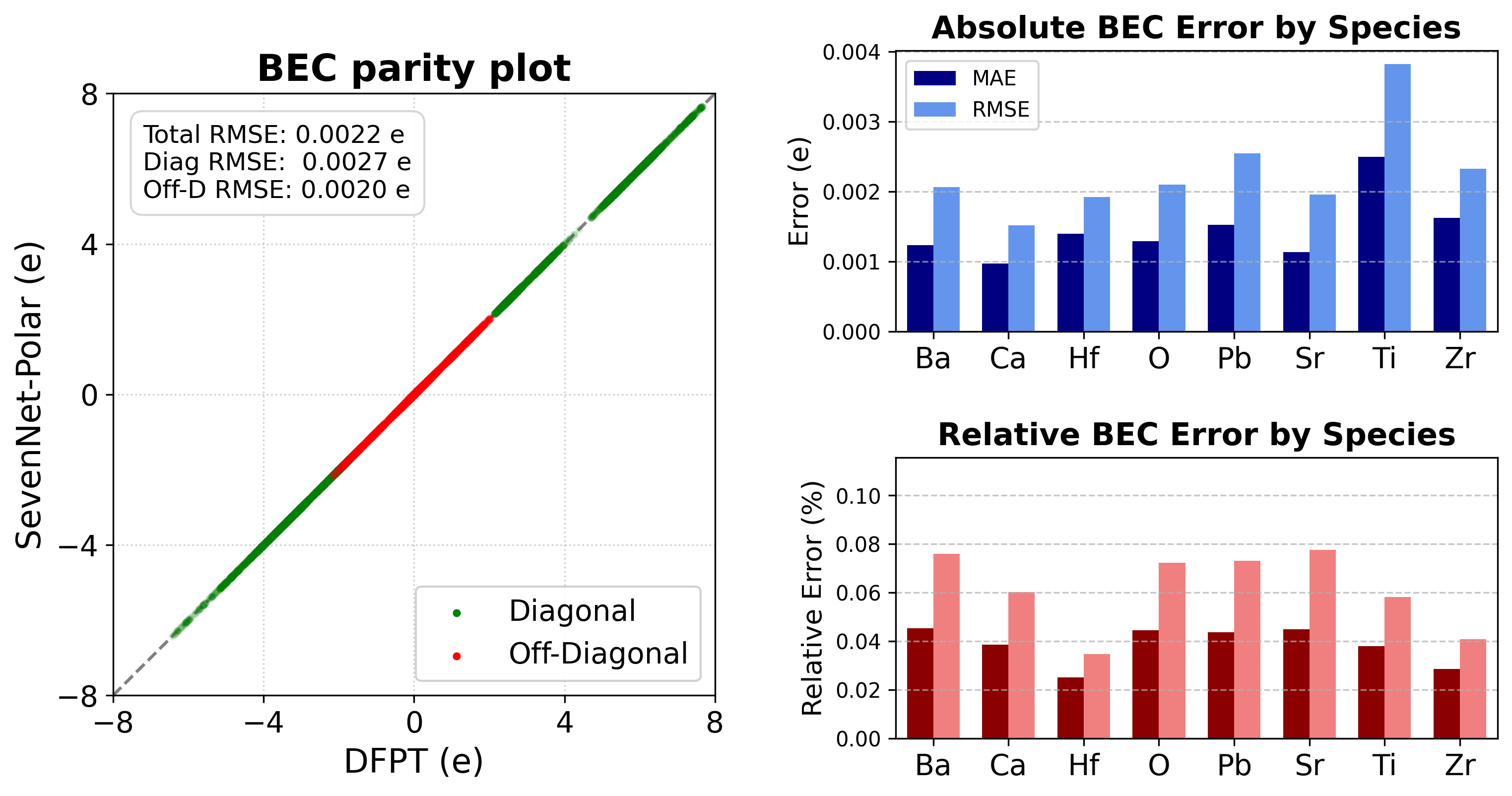}
    \caption{Evaluation of SevenNet-PS-M on the perovskite data set}
    \label{fig:specialized_on_perovskite_test}
\end{figure}

\autoref{fig:specialized_on_zro2_test}, \autoref{fig:specialized_on_li3po4_test} and \autoref{fig:specialized_on_perovskite_test} show the parity plots and errors for SevenNet-PS-M on each separate data set.

\clearpage

\section{Simulation time added by the BEC prediction}\label{sec:bec_overhead}
    
\begin{table}[htbp]
\centering
\caption{\label{tab:bec_overhead} Computational overhead of predicting Born effective charges on a single NVIDIA GeForce RTX 4090. The evaluation compares the standard \texttt{e3nn} against the optimized \texttt{FlashTP} implementation for SevenNet-PM-M. A regular SevenNet with identical parameters was trained to predict only energy, forces, and stress, while SevenNet-PM-M includes the BEC tensor prediction in addition to these three quantities.}
\begin{tabular}{@{}lcccc@{}}
\toprule
Execution Mode & System Size & SevenNet & SevenNet-Polar & Overhead \\
 & (Atoms) & (ms/step) & (ms/step) & (\%) \\
\midrule
Standard & 96   & 81.45  & 170.11 & +108.9 \\
Standard & 216  & 168.55 & 380.09 & +125.5 \\
Standard & 324  & 231.00 & 568.51 & +146.1 \\
Standard & 432  & 294.80 & 758.18 & +157.2 \\
\midrule
\texttt{FlashTP}  & 96   & 52.65  & 27.86  & -47.1 \\
\texttt{FlashTP}  & 216  & 56.73  & 33.87  & -40.3 \\
\texttt{FlashTP}  & 324  & 63.14  & 41.94  & -33.6 \\
\texttt{FlashTP}  & 432  & 85.34  & 53.27  & -37.6 \\
\texttt{FlashTP}  & 576  & 83.91  & 68.57  & -18.3 \\
\texttt{FlashTP}  & 768  & 80.54  & 89.52  & +11.1 \\
\texttt{FlashTP}  & 960  & 97.24  & 110.05 & +13.2 \\
\texttt{FlashTP}  & 1200 & 112.64 & 134.90 & +19.8 \\
\texttt{FlashTP}  & 1500 & 137.04 & 165.94 & +21.1 \\
\texttt{FlashTP}  & 1800 & 139.60 & 197.62 & +41.6 \\
\texttt{FlashTP}  & 2160 & 182.28 & 235.51 & +29.2 \\
\texttt{FlashTP}  & 2592 & 212.58 & 282.67 & +33.0 \\
\texttt{FlashTP}  & 3024 & 263.06 & 334.59 & +27.2 \\
\texttt{FlashTP}  & 3528 & 313.04 & 395.88 & +26.5 \\
\bottomrule
\end{tabular}
\end{table}
    
To quantify the computational penalty associated with evaluating the Born effective charge (BEC) tensor, we benchmarked the execution time per molecular dynamics step with and without BEC prediction (\autoref{tab:bec_overhead}). The baseline evaluation (without BEC prediction) is done with standard SevenNet. Without \texttt{FlashTP}, adding the prediction of the BEC tensor more than doubles the simulation time (+146.1\% overhead with 324 atoms). However, the introduction of the \texttt{FlashTP} backend largely mitigates this overhead, as the computational penalty for predicting full rank-2 BEC tensors is restricted to at most 41.6\% and stabilizes around 30 \% for systems with more than 2,000 atoms. Remarkably, this algorithmic efficiency persists even when scaling to supercells exceeding 3,000 atoms, demonstrating that molecular dynamics under an electric field can now be achieved on a single consumer GPU with minimal performance degradation.
    
\clearpage
\section{Atom-resolved BEC prediction along NEB trajectory}\label{sec:neb_supplementary}

\begin{figure}[htbp]
    \centering
    \includegraphics[width=\linewidth]{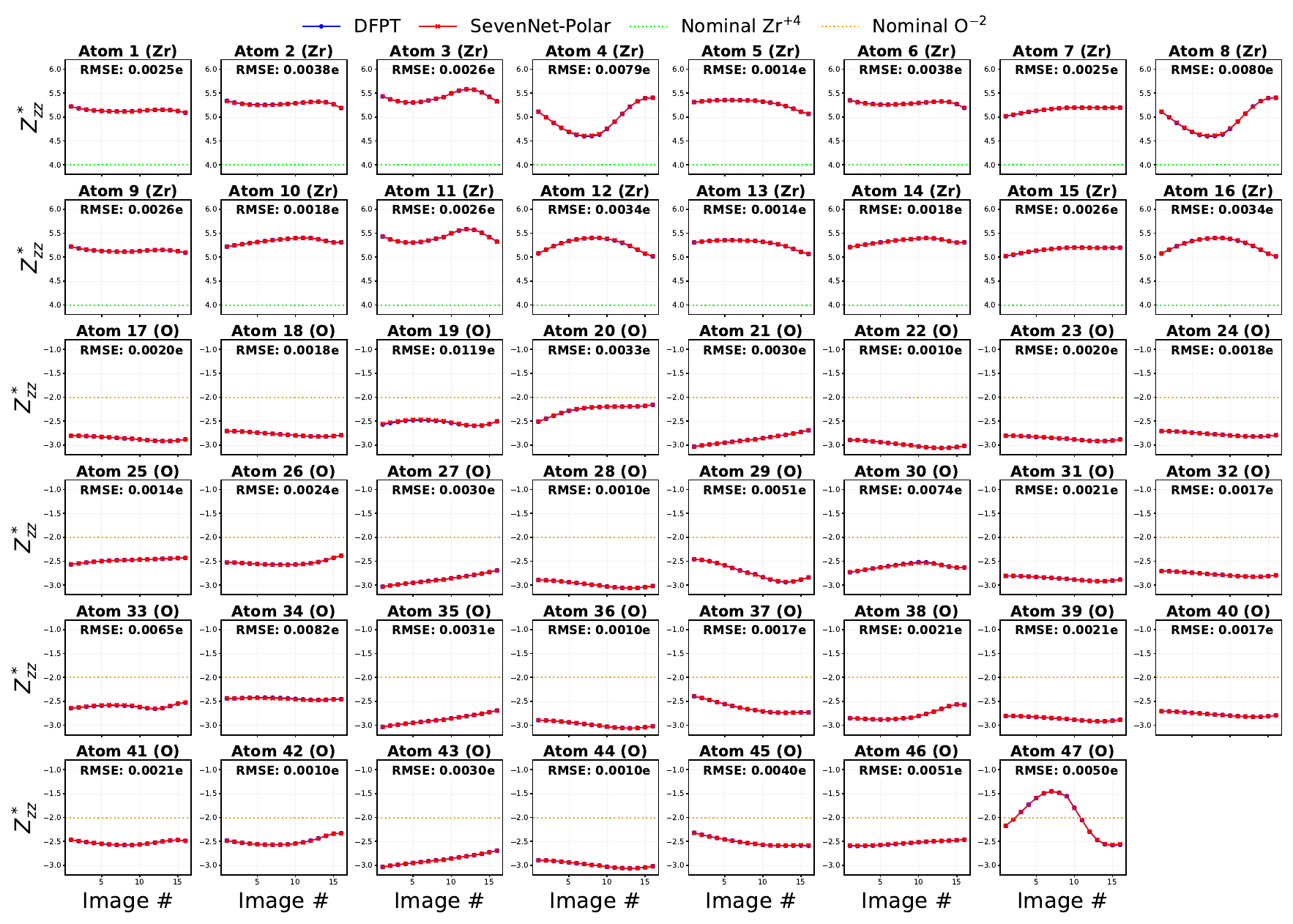}
    \caption{Evaluation of the $Z^*_{zz}$ component of the BEC tensor for each individual atom along a NEB trajectory, using DFT-calculated images. The nominal charges are indicated by green (\ce{Zr^{+4}}) and orange (\ce{O^{-2}}) dotted lines.}
    \label{fig:neb_004_sevennet_vs_nominal_6x8}
\end{figure}

The $Z^*_{zz}$ component of the Born effective charge tensor predicted by the SevenNet-PM-M is shown in \autoref{fig:neb_004_sevennet_vs_nominal_6x8}.

\clearpage
\section{Atom-resolved BEC prediction in $\Sigma 5(310)/[001]$ grain boundary in cubic-\ce{ZrO2}}\label{sec:gb1_supplementary}
    
\begin{figure}[htbp]
    \centering
    \includegraphics[width=1.0\linewidth]{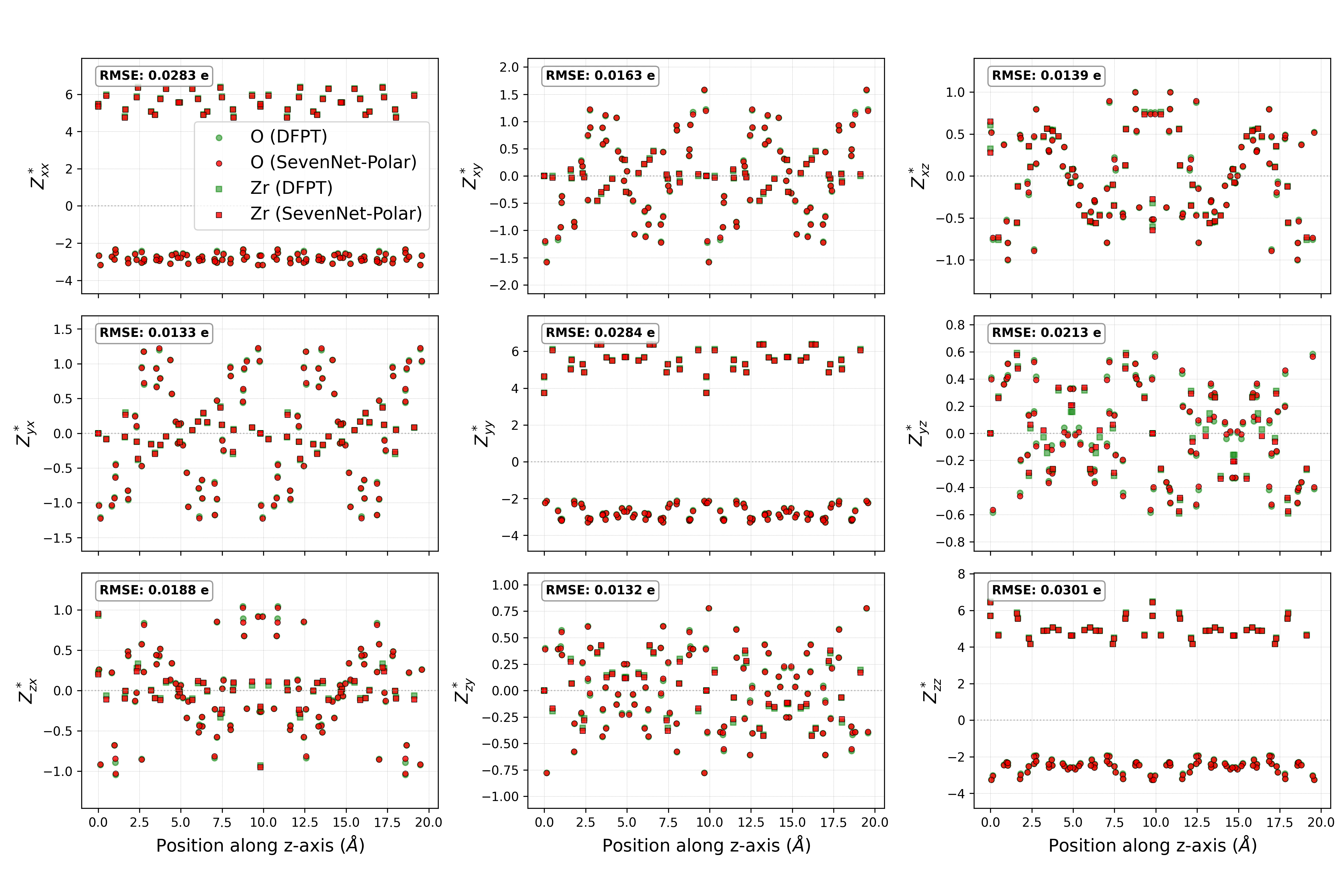}
    \caption{Detailed results of the prediction of the BEC tensor on a $\Sigma 5(310)/[001]$ model from Ref.\cite{Arai_2025}. Each subplot shows a single BEC component, and each point is the prediction for a single ion. Squares represent Zr ions while circles represent O ions. DFPT results are shown in red while SevenNet-Polar (SevenNet-PM-M) results are shown in blue.}
    \label{fig:GB_1_bec_z_comparison}
\end{figure}
    
\autoref{fig:GB_1_bec_z_comparison} shows the BEC tensor prediction for each atom, component by component. The predicted values are close to the DFPT results, with each component having RMSE of 0.03 e or below.

\clearpage

\section{Scaling law on \ce{SiO2} and \ce{BaTiO3} data sets}\label{sec:scaling_law_sio2_batio3}

\begin{figure*}[htbp]
    \centering
    \includegraphics[width=\linewidth]{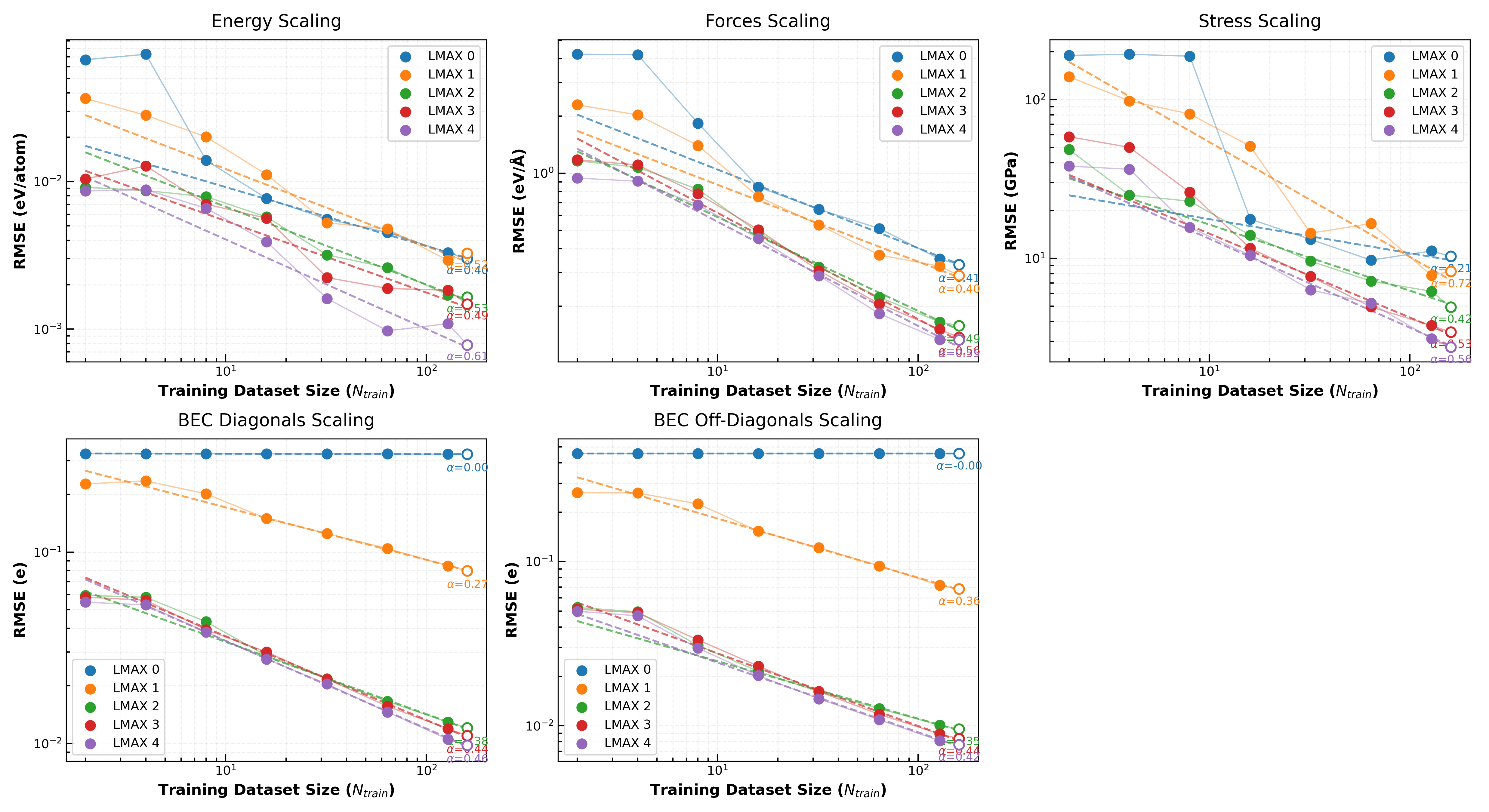}
    \caption{SevenNet-Polar multitask (SevenNet-PM family) RMSE power law for the \ce{SiO2} data set \cite{Schmiedmayer_2026_data}}
    \label{fig:sio2_multitask_power_law}
\end{figure*}

\begin{figure*}[htbp]
    \centering
    \includegraphics[width=\linewidth]{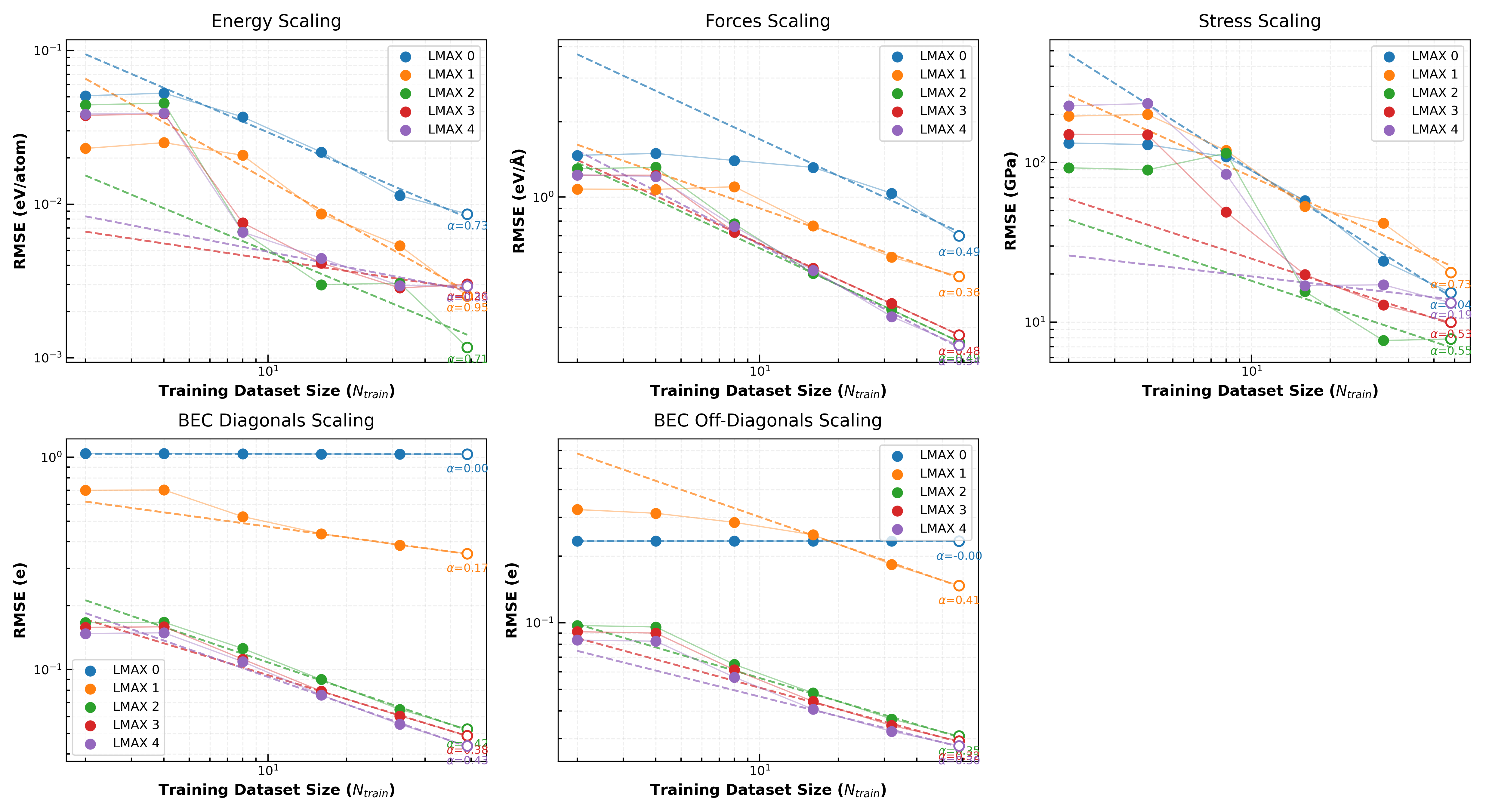}
    \caption{SevenNet-Polar multitask (SevenNet-PM family) RMSE power law for the \ce{BaTiO3} data set \cite{Schmiedmayer_2026_data}}
    \label{fig:batio3_multitask_power_law}
\end{figure*}

We evaluated the scaling laws for the RMSE using the data set by Schmiedmayer et al. \cite{Schmiedmayer_2026_data}. While the data set has relatively few structures, the power laws for the error on each quantity can clearly be identified for both \ce{SiO2} in \autoref{fig:sio2_multitask_power_law} and \ce{BaTiO3} in \autoref{fig:batio3_multitask_power_law}. As the RMSE curves have not yet reached the saturation regime, obtaining more data points to enrich these data sets would undoubtedly have a positive impact on the accuracy of future models.

\clearpage
\section{Detailed performance analysis with \texttt{LAMMPS} }\label{sec:detailed_benchmark_gpu}
    
The increase in maximum system size enabled by the lower memory footprint of \texttt{FlashTP} is shown in \autoref{tab:max_system_size} for an extensive selection of GPUs. 
        
\begin{table*}[htpb]
    \centering
    \caption{Maximum system size capacity, reported as number of atoms, before encountering Out-Of-Memory (OOM) errors across different hardware configurations. Results are shown as \texttt{FlashTP} OFF $\rightarrow$ \texttt{FlashTP} ON, with the corresponding capacity multiplier in parentheses.}
    \label{tab:max_system_size}
    \resizebox{\textwidth}{!}{%
    \begin{tabular}{@{}lcccc@{}}
    \toprule
    \textbf{Hardware Configuration} &
    \textbf{SevenNet-Omni-12i} &
    \textbf{SevenNet-0} &
    \textbf{SevenNet-PM-M} &
    \textbf{SevenNet-PM-S} \\
    \midrule
    \texttt{A100 (80GB) -- serial} &
    1,200 $\rightarrow$ 9,720 ($8.1\times$) &
    9,720 $\rightarrow$ 86,640 ($8.9\times$) &
    768 $\rightarrow$ 12,000 ($15.6\times$) &
    5,376 $\rightarrow$ 46,080 ($8.6\times$) \\
    
    \texttt{A100 (80GB) x 1 -- parallel} &
    1,200 $\rightarrow$ 9,720 ($8.1\times$) &
    14,520 $\rightarrow$ 82,308 ($5.7\times$) &
    960 $\rightarrow$ 12,000 ($12.5\times$) &
    7,776 $\rightarrow$ 28,392 ($3.7\times$) \\
    
    \texttt{A100 (80GB) x 2 -- parallel} &
    1,800 $\rightarrow$ 1,800 ($1.0\times$) &
    28,392 $\rightarrow$ 111,132 ($3.9\times$) &
    2,592 $\rightarrow$ 24,336 ($9.4\times$) &
    14,520 $\rightarrow$ 100,800 ($6.9\times$) \\
    
    \texttt{RTX 4090 (24GB) -- serial} &
    324 $\rightarrow$ 3,024 ($9.3\times$) &
    4,116 $\rightarrow$ 24,336 ($5.9\times$) &
    324 $\rightarrow$ 4,116 ($12.7\times$) &
    2,160 $\rightarrow$ 15,972 ($7.4\times$) \\
    
    \texttt{RTX 6000 (96GB) -- serial} &
    1,500 $\rightarrow$ 12,000 ($8.0\times$) &
    15,972 $\rightarrow$ 100,800 ($6.3\times$) &
    1,500 $\rightarrow$ 15,972 ($10.6\times$) &
    9,720 $\rightarrow$ 69,984 ($7.2\times$) \\   
    
    \midrule
    
    \texttt{A100 (40GB) -- serial} &
    576 $\rightarrow$ 5,376 ($9.3\times$) &
    6,912 $\rightarrow$ 40,500 ($5.9\times$) &
    576 $\rightarrow$ 6,912 ($12.0\times$) &
    4,116 $\rightarrow$ 28,392 ($6.9\times$) \\
    
    \texttt{A100 (40GB) x 1 -- parallel} &
    576 $\rightarrow$ 4,704 ($8.2\times$) &
    6,912 $\rightarrow$ 40,500 ($5.9\times$) &
    576 $\rightarrow$ 6,912 ($12.0\times$) &
    4,116 $\rightarrow$ 28,392 ($6.9\times$) \\
    
    \texttt{A100 (40GB) x 2 -- parallel} &
    1,200 $\rightarrow$ 9,720 ($8.1\times$) &
    14,520 $\rightarrow$ 77,976 ($5.4\times$) &
    1,200 $\rightarrow$ 9,720 ($8.1\times$) &
    7,776 $\rightarrow$ 49,152 ($6.3\times$) \\
    
    \texttt{A100 (40GB) x 4 -- parallel} &
    2,160 $\rightarrow$ 19,008 ($8.8\times$) &
    28,392 $\rightarrow$ 158,976 ($5.6\times$) &
    2,592 $\rightarrow$ 24,336 ($9.4\times$) &
    14,520 $\rightarrow$ 100,800 ($6.9\times$) \\
    
    \texttt{A100 (40GB) x 8  -- parallel} &
    4,704 $\rightarrow$ 40,500 ($8.6\times$) &
    55,488 $\rightarrow$ 302,760 ($5.5\times$) &
    5,376 $\rightarrow$ 49,152 ($9.1\times$) &
    30,576 $\rightarrow$ 172,800 ($5.7\times$) \\
    
    \texttt{A100 (40GB) x 16 -- parallel} &
    7,776 $\rightarrow$ 77,976 ($10.0\times$) &
    111,132 $\rightarrow$ 607,836 ($5.5\times$) &
    9,720 $\rightarrow$ 96,000 ($9.9\times$) &
    62,424 $\rightarrow$ 393,216 ($6.3\times$) \\
    
    \texttt{A100 (40GB) x 32 -- parallel} &
    19,008 $\rightarrow$ 158,976 ($8.4\times$) &
    227,448 $\rightarrow$ 1,245,876 ($5.5\times$) &
    20,736 $\rightarrow$ 172,800 ($8.3\times$) &
    121,968 $\rightarrow$ 675,792 ($5.5\times$) \\
    
    \texttt{A100 (40GB) x 64 -- parallel} &
    35,280 $\rightarrow$ 302,760 ($8.6\times$) &
    444,312 $\rightarrow$ 2,506,320 ($5.6\times$) &
    40,500 $\rightarrow$ 345,960 ($8.5\times$) &
    227,448 $\rightarrow$ 1,500,000 ($6.6\times$) \\
    
    \bottomrule
    \end{tabular}%
    }
\end{table*}
            
\clearpage            
\section{Simulation speed on a consumer-grade GPU}

\begin{figure}[htbp]
    \centering
    \includegraphics[width=0.8\linewidth]{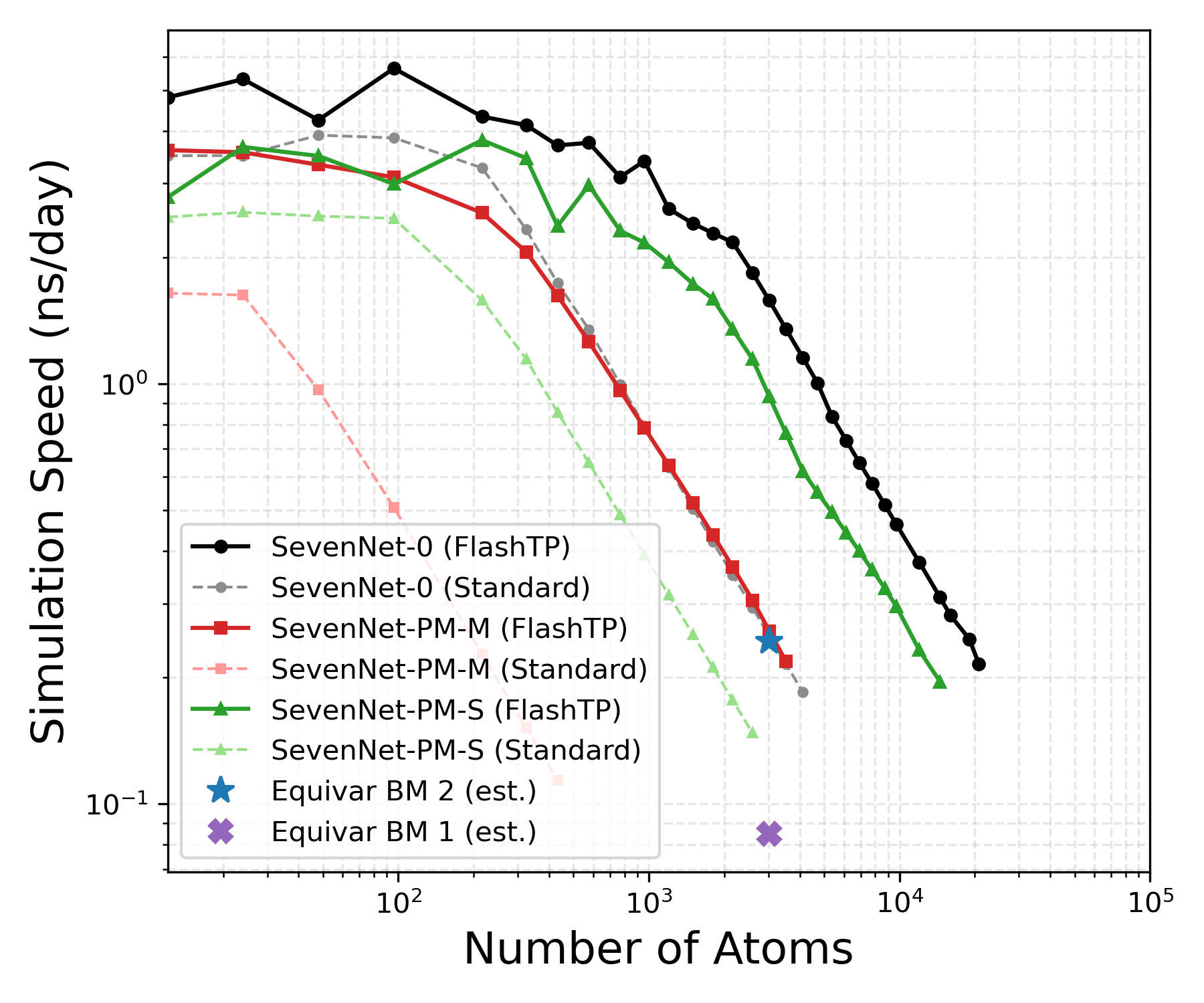}
    \caption{Simulation speed with respect to the number of atoms, tested on an NVIDIA GeForce RTX 4090 GPU. The time was averaged over 100 MD steps using ASE. Estimated reference points for \texttt{equivar} BM 1 and BM 2 at 3,024 atoms are included for comparison, assuming that the simulation time is entirely driven by the time for the BEC predictions.} \label{fig:Speed_vs_Atoms_4090}
\end{figure}

\autoref{fig:Speed_vs_Atoms_4090} shows the simulation speed as a function of the number of atoms for a consumer-grade NVIDIA GeForce RTX 4090 GPU.

\clearpage
\section{Data set format}\label{sec:ExtxyzBEC}
    
To store atomic configurations including the information about BEC, we use extended XYZ (\texttt{extxyz}) files. We append dedicated columns for each of the 9 Cartesian components of the Born effective charge tensor. This standardizes the data set and enables direct parsing using the Atomic Simulation Environment (ASE)~\cite{Larsen_2017}. 
    
The structure of the \texttt{extxyz} format used in our training pipeline is illustrated below. Note the \texttt{born\_effective\_charges:R:9} declaration in the \texttt{Properties} key, which dictates the shape of the tensor array:
    
\begin{table}[htbp]
\begin{lstlisting}[breaklines=true, basicstyle=\scriptsize\ttfamily, frame=lines]
48
Lattice="7.43337441 0.0 0.0 0.0 7.43337441 0.0 0.0 0.0 10.84170227" Properties=species:S:1:pos:R:3:born_effective_charges:R:9:forces:R:3 energy=-449.9320341 stress="..." free_energy=-449.9320341 pbc="T T T"
Zr  0.10822830  7.40199494  0.10891065  5.73233842  0.04824380 -0.75964342  0.07275877  5.24113599  0.07192443 -0.69945296 -0.12277566  5.09081450 -0.46672206 -1.41394895  0.74414710
Zr  0.03824062  0.03993298  5.42354573  5.82244994  0.61572568  0.13617175  0.35999269  5.11591290 -0.20195818 -0.04126178  0.15551198  5.23369038  0.21714642 -0.20947993 -0.53408527
...
O   1.93781500  0.20044168  1.27776996 -2.91532665  0.10408339  0.09433348 -0.08518494 -2.43950092 -0.57239810  0.17061071 -0.30820757 -2.58964010 -0.49857488 -1.20356798 -0.90028188
O   1.80263507  0.22060092  6.49205881 -3.46886588 -0.12728852 -0.33323842 -0.10410875 -2.00869952 -0.06279408 -0.27883360 -0.24865164 -2.80402712 -0.22375810 -0.90747978  0.24495925
...
\end{lstlisting}
\label{tab:extxyz_format}
\caption{Truncated example of the \texttt{extxyz} file format used to store configurations, scalar energies, vectorial forces, and the rank-2 Born effective charge tensors for \ce{ZrO2}.}
\end{table}

\clearpage
\section{Modifications of the SevenNet source code} \label{sec:CodeChanges}
    
Extending the SevenNet architecture to support Born effective charge (BEC) prediction required targeted modifications to the data pipeline and the model's objective function. Firstly, the data set parser was updated to extract and format the ground-truth BEC targets as $N \times 3 \times 3$ Cartesian arrays, where $N$ is the number of atoms in the configuration. Secondly, a custom \texttt{BECLoss} class was implemented to seamlessly integrate the tensor error into the global multitask loss function. Finally, rigorous mathematical projections were introduced to handle the conversion between the standard Cartesian basis and the $E(3)$-equivariant irreducible representations (irreps) required by the network, ensuring that rotational symmetries are exactly preserved during training. In addition, we wrote specific classes to calculate the RMSE error on BEC, separating diagonal components from off-diagonal ones. In nominal charge models, off-diagonal contributions are assumed to be zero. 
    
The following files within the SevenNet code-base were modified:
\begin{itemize}
    \item \textbf{\texttt{sevenn/\_keys.py}} and \textbf{\texttt{sevenn/\_const.py}}: Added configuration constants, dictionary keys, and default metric registries to support BEC training flags and error tracking.
    \item \textbf{\texttt{sevenn/train/dataload.py}} and \textbf{\texttt{sevenn/train/graph\_dataset.py}}: Updated the data parsing pipeline to extract and correctly format the multi-dimensional $N \times 3 \times 3$ tensorial targets directly from extended XYZ files.
    \item \textbf{\texttt{sevenn/model\_build.py}}: Modified the graph construction logic to enforce higher-order equivariant features ($l_{max} \ge 2$, full parity) and appended the tensorial \texttt{IrrepsLinear} readout head.
    \item \textbf{\texttt{sevenn/train/loss.py}}: Introduced the \texttt{BECLoss} class to handle mathematically exact Cartesian-to-irrep projections and to correct gradient scaling artifacts common in tensor learning.
    \item \textbf{\texttt{sevenn/calculator.py}}: Extended the \texttt{ASE} calculator interface to mathematically reconstruct and output standard Cartesian BEC tensors during inference.
    \item \textbf{\texttt{sevenn/error\_recorder.py}}: Created custom diagnostic metrics to isolate and track the physical Root Mean Square Error (RMSE) for the diagonal and off-diagonal tensor components independently.
\end{itemize}
    
\textit{Remark}: The BEC loss had to be multiplied by 9 to account for proper weighting due to the way RMSE is calculated after flattening the tensor. In standard PyTorch implementations, flattening a $[N_{atoms}, 9]$ tensor prior to a mean-reduction loss operation averages the squared error over $N_{atoms} \times 9$ elements. This artificially suppresses the gradient magnitude of the BEC loss by an order of magnitude relative to the per-atom energy loss. To correct this dimensional artifact, we implement a normalized per-atom BEC loss $\mathcal{L}_{\text{BEC}}$:
\begin{equation}
    \mathcal{L}_{\text{BEC}} = \frac{9}{N_{atoms}} \sum_{i=1}^{N_{atoms}} \frac{1}{9} \sum_{c=1}^{9} \left( \hat{Y}_{i,c} - Y_{i,c} \right)^2
\end{equation}
By explicitly multiplying the flattened mean loss by $9$, we restore the gradient scale, ensuring that the optimizer dedicates proportionate capacity to learning the BEC features without requiring manual user tuning of hyperparameter loss weights.
    
\clearpage
\section{Added keywords for the interface of SevenNet} \label{sec:NewKeywords}
    
To integrate the Born effective charge (BEC) training into the existing SevenNet framework, the standard YAML configuration file (\texttt{input.yaml}) was extended. Four new parameters were introduced to control the activation, weighting, and monitoring of the BEC tensorial predictions:
    
\begin{enumerate}
    \item \texttt{is\_train\_bec: True} (under the \texttt{train} block): 
    This boolean flag acts as the primary toggle for the BEC module. When set to \texttt{True}, it activates the equivariant tensorial readout head within the neural network architecture, instructing the model to output an $N \times 3 \times 3$ tensor. It also triggers the data loader to fetch the \texttt{born\_effective\_charges} arrays from the extended XYZ training files.
    
    \item \texttt{bec\_loss\_weight: 10.0} (under the \texttt{train} block): 
    This scalar defines the relative penalty of the BEC prediction error within the global loss function, alongside \texttt{energy\_loss\_weight}, \texttt{force\_loss\_weight}, and \texttt{stress\_loss\_weight}. Because the numerical values of the BECs (in units of elementary charge $e$) often exhibit different variance scales compared to energies (eV) or forces (eV/\AA), this weight must be tuned to prevent the BEC gradients from either dominating or being overwhelmed during the backpropagation process.
    
    \item \texttt{['BornEffectiveCharges', 'DiagRMSE']} (under \texttt{error\_record}): 
    A new logging metric that isolates and tracks the Root Mean Square Error (RMSE) specifically for the diagonal components of the BEC tensors ($Z^*_{xx}, Z^*_{yy}, Z^*_{zz}$). Since diagonal components strictly dictate the primary polarization response and are typically an order of magnitude larger than off-diagonal terms, isolating this metric is crucial for diagnosing the physical accuracy of the model.
    
    \item \texttt{['BornEffectiveCharges', 'OffDiagRMSE']} (under \texttt{error\_record}): 
    A new complementary logging metric that tracks the RMSE of the off-diagonal components ($Z^*_{xy}, Z^*_{xz}, Z^*_{yx}, Z^*_{yz}, Z^*_{zx}, Z^*_{zy}$). Monitoring this error separately ensures that the neural network correctly captures the subtle spatial asymmetries and local shear distortions of the dielectric response, which are particularly important in highly defective or strained systems.
\end{enumerate}
    
This minimizes the amount of modifications needed to set SevenNet to train on BEC data, enabling users to write their code or use \texttt{sevenn} with minimal configuration changes to train their model.

\clearpage
\section{\texttt{ASE} Calculator and \texttt{LAMMPS} Interface}\label{sec:calculators_and_lammps}

The \texttt{SevenNet} package provides an Atomic Simulation Environment (ASE) calculator and a \texttt{LAMMPS} interface to facilitate atomistic simulations, including the prediction of Born effective charges (BEC). Both interfaces are optimized for \texttt{PyTorch}, natively supporting hardware acceleration via single and multi-GPU configurations, as well as the \texttt{FlashTP} library for efficient tensor product evaluations.

\subsection{\texttt{ASE} Calculator}
    
The \texttt{ASE} calculator for \texttt{SevenNet} mirrors the standard usage typical of machine learning potentials in the \texttt{ASE} framework, while extending its capabilities to predict tensorial properties such as the Born effective charges. 
    
To initialize the calculator, users instantiate the \texttt{SevenNetCalculator} by providing the model checkpoint. Once the calculator is attached to an \texttt{ASE} \texttt{Atoms} object, invoking standard commands (e.g., \texttt{get\_forces()}, \texttt{get\_potential\_energy()}) will trigger the model inference.
    
The Born effective charge prediction is enabled automatically if the loaded model supports it. The BEC tensor can be retrieved from the calculator's results dictionary as a $3 \times 3$ Cartesian tensor per atom. Internally, the network's output—represented as 9-component \texttt{e3nn} irreducible representations ($1\times0e + 1\times1e + 1\times2e$)—is seamlessly converted back to the Cartesian basis.
    
A brief example of extracting the BEC tensor in \texttt{ASE} is shown below:
\begin{verbatim}
from ase.io import read
from sevenn.calculator import SevenNetCalculator
    
# Load structure, attach calculator
# and evaluate energy, forces, and BEC 
atoms = read("structure.xyz")
atoms.calc = SevenNetCalculator(model="checkpoint.pth")
atoms.get_potential_energy()
bec_tensor = atoms.calc.results['born_effective_charges']
\end{verbatim}
    
\subsection{\texttt{LAMMPS} Interface}
For large-scale molecular dynamics and multi-GPU acceleration, \texttt{SevenNet} provides a deeply integrated \texttt{LAMMPS} interface. Built on PyTorch's C++ API (\texttt{libtorch}), the \texttt{pair\_style e3gnn} implementation ensures high-performance inference and natively supports both \texttt{FlashTP} for tensor product optimizations and multi-GPU distribution via MPI and \texttt{LAMMPS} spatial decomposition.
    
When running molecular dynamics with models trained to predict Born effective charges, the \texttt{LAMMPS} interface can couple the predicted BEC tensors with an externally applied electric field to calculate the resulting electrostatic forces on each atom. The applied electric field vector can be provided directly within the \texttt{LAMMPS} input script.
    
The following example demonstrates how to set up the \texttt{pair\_style} for \texttt{SevenNet} and apply an electric field of 0.02 V/\AA\ along the $z$-direction. The corresponding forces generated by the coupling of the BEC and the electric field are automatically added to the total force.

\subsubsection{Serial}
    
\begin{verbatim}
pair_style      e3gnn
# Model path, applied electric field (ex ey ez), and chemical species
pair_coeff      * * deployed_model.pt efield 0.0 0.0 0.02 Zr O
\end{verbatim}
            
\subsubsection{Parallel}

\begin{verbatim}
pair_style      e3gnn/parallel
# Model path, number of layers, # applied electric field (ex ey ez),
# and chemical species
pair_coeff * * 4 deployed_model_parallel.pt efield 0.0 0.0 0.02 Zr O
\end{verbatim}

\clearpage


\end{document}